%% file: ms.tex
\documentclass[usenatbib,usegraphicx]{mn2e}
\usepackage{amssymb}
\usepackage[fleqn]{amsmath}
\usepackage{charter}
\usepackage{mathdesign}
\usepackage{url}
\usepackage{color}
\providecommand{\adsurl}[1]{\href{#1}{ADS}}

\newcommand{\oA}{\omega_\mathrm{A}}
\newcommand{\imag}{{\rm i}}
\newcommand{\ep}{\varepsilon}

\pagerange{\pageref{firstpage}--\pageref{lastpage}} \pubyear{2015}

\def\LaTeX{L\kern-.36em\raise.3ex\hbox{a}\kern-.15em
T\kern-.1667em\lower.7ex\hbox{E}\kern-.125emX}

\input shortcuts

\title[Protoplanetary disc accretion]{Magnetically driven accretion in protoplanetary discs}

\author[Simon et al.]
{Jacob B.~Simon$^{1,2,3 \ast}$,
Geoffroy Lesur$^{4,5 \dagger}$,
Matthew W.~Kunz$^{6 \star}$,
Philip J.~Armitage$^{3,7 \ddagger}$
\\$^1$Department of Space Studies,
Southwest Research Institute, Boulder, CO 80302, USA
\\$^2$Sagan Fellow
\\$^3$JILA,
University of Colorado \& NIST,
440 UCB,
Boulder, CO 80309-0440, USA
\\$^4$Univ.~Grenoble Alpes, IPAG, 38000 Grenoble, France
\\$^5$CNRS, IPAG, 38000 Grenoble, France
\\$^6$Department of Astrophysical Sciences, 4 Ivy Lane, Peyton Hall, Princeton University, Princeton NJ 08544, USA
\\$^7$Department of Astrophysical and Planetary Sciences, 
University of Colorado at Boulder, Boulder, CO 80309, USA
\\Email: $^\ast$jbsimon.astro@gmail.com,
$^\star$mkunz@princeton.edu,
$^\dagger$geoffroy.lesur@obs.ujf-grenoble.fr, 
$^\ddagger$pja@jilau1.colorado.edu}

\begin{document}

\label{firstpage}

\maketitle

\begin{abstract} 
We characterize magnetically driven accretion at radii between $1~{\rm au}$ and $100~{\rm au}$ in protoplanetary discs, using a series of local non-ideal magnetohydrodynamic (MHD) simulations. The simulations assume a Minimum Mass Solar Nebula (MMSN) disc that is threaded by a net vertical magnetic field of specified strength. Confirming previous results, we find that the Hall effect has only a modest impact on accretion at $30~{\rm au}$, and essentially none at $100~{\rm au}$. At $1$--$10~{\rm au}$ the Hall effect introduces a pronounced bi-modality in the accretion process, with 
vertical magnetic fields aligned to the disc rotation supporting a strong laminar Maxwell stress that is absent if the field is anti-aligned. In the anti-aligned case, we instead find evidence for bursts of turbulent stress at $5$--$10~{\rm au}$, which we tentatively identify with the non-axisymmetric Hall-shear instability. The presence or absence of these bursts depends upon the details of the adopted chemical model, which suggests that appreciable regions of actual protoplanetary discs might lie close to the borderline between laminar and turbulent behaviour. Given the number of important control parameters that have already been identified in MHD models, quantitative predictions for disc structure in terms of only radius and accretion rate appear to be difficult. Instead, we identify robust qualitative tests of magnetically driven accretion. These include the presence of turbulence in the outer disc, independent of the orientation of the vertical magnetic fields,  and a Hall-mediated bi-modality in turbulent properties extending from the region of thermal ionization to $10~{\rm au}$.
\end{abstract}

\begin{keywords}
{accretion, accretion discs  -- protoplanetary discs -- MHD -- instabilities -- turbulence}
\end{keywords}

\section{Introduction}

Gas in protoplanetary discs loses angular momentum and accretes onto the central protostar at rates far greater than can be explained as a consequence of molecular viscosity \citep{hartmann98a}. While myriad physical processes that might explain the observed rates continue to be debated \citep[see the reviews by][]{armitage11,turner14}, the magnetorotational instability (MRI; \citealt{balbus98}) is widely considered to be the leading contender for providing the necessary enhanced angular-momentum transport. That this is true not just for protoplanetary discs, but also for a wide variety of accreting systems (e.g.~compact objects), attests to the fact that the basic prerequisites for the instability are quite generic: a negative angular velocity gradient, a sub-thermal magnetic field, and enough free charges to provide sufficient coupling between the gas and the magnetic field. What makes protoplanetary discs particularly interesting from the perspective of the MRI is this final requirement; typical degrees of ionization in the deep interiors of such discs range from $\sim$$10^{-9}$ to as low as $\sim$$10^{-17}$, and it is presently unclear how resilient the MRI is under these conditions.

One of the first models of MRI-driven accretion in protoplanetary discs was constructed by \cite{gammie96}, who proposed that Ohmic resistivity (due to weak coupling of charged species to the magnetic field) quenches the MRI in a `dead' mid-plane region surrounded by `active layers', which are ionized by cosmic rays. This layered-accretion scenario has been refined as additional non-thermal ionization sources (X-rays, UV radiation) and {\it non-ideal} magnetohydrodynamic (MHD) effects (ambipolar diffusion and the Hall effect) have been included \citep[e.g.][]{igea99,sano00,fromang02,salmeron03,salmeron05,salmeron08,ilgner06,bai09,bai11a,wardle12}.  Several theoretical studies, spanning both analytical \citep[e.g.][]{blaes94,kunz04,desch04} and numerical work \citep[e.g.][]{hawley98,bai11a,bai13b,simon13a,simon13b}, have shown that ambipolar diffusion plays a major role in determining the level of turbulence in protoplanetary discs. In the outer regions (radii $R \gtrsim 30~{\rm au}$), ambipolar diffusion acts to reduce the strength of MRI-driven turbulence close to the disc mid-plane; turbulence is stronger above this `ambipolar damping zone' in a thin layer of gas strongly ionized by FUV photons \citep{perez-becker11b,simon13a,simon13b}. To produce accretion rates consistent with observations, a large-scale magnetic field threading the disc perpendicular to the disc plane is required \citep{simon13a,simon13b}. This net field enhances turbulence, despite the damping effect of ambipolar diffusion, and allows additional angular-momentum loss via a magnetic wind (akin to the \citealt{blandford82} mechanism). In the inner disc, both local \citep{bai13b} and global \citep{gressel15} simulations have shown that ambipolar diffusion can quench the MRI in the purported active layers of Gammie's (1996\nocite{gammie96}) model. In those cases, a large-scale magnetic field is again necessary to drive accretion at the observationally inferred rates, this time solely via a magnetic wind in a manner akin to previous models of wind-driven accretion in protoplanetary discs \cite[e.g.][]{wardle93,shu94,salmeron07}.

The influence of the Hall effect has been considered in several studies \cite[e.g.][]{wardle99a,balbus01,sano02a,sano02b}, but only recently has it been possible to numerically investigate the case in which the Hall term dominates over both the inductive term {\em and} the other non-ideal effects. The Hall effect operates at densities intermediate to those where Ohmic diffusion (high densities) and ambipolar diffusion (low densities) dominate \citep[e.g.][]{wardle07}. Where dominant, the Hall effect leads to a qualitatively different mode of angular-momentum transport \citep{kunz13,lesur14,bai14a}, and the nature of the associated stress depends on the {\em orientation} of the large-scale magnetic field with respect to the angular momentum of the disc \citep[as suggested by the linear analysis of][]{wardle99a}. This is quantified by the sign of $\bm{\Omega}\bcdot\bm{B}$ ($\bm{\Omega}$ is the angular velocity vector and $\bm{B}$ is the large scale magnetic field). For $\bm{\Omega}\bcdot\bm{B} < 0$, the Hall effect was found to reduce turbulent angular-momentum transport significantly; for $\bm{\Omega}\bcdot\bm{B} > 0$, the transport was found to be enhanced, occurring primarily through large-scale laminar stresses and magnetic winds.
 
In this paper, we further examine the role of magnetic fields and non-ideal MHD in driving accretion in protoplanetary discs. While we include all three non-ideal effects -- Ohmic dissipation, ambipolar diffusion, and the Hall effect -- our main focus is on the influence of the latter. We first show that the Hall effect has minimal impact on the behaviour of the accretion flow in the outer disc. This confirms predictions based upon linear analyses as well as the recent numerical results of \cite{bai15}, who used a different numerical scheme that the one we employ herein. At smaller radii in discs with $\bmath{\Omega}\cdot\bmath{B} <0$, we identify a new regime of `bursty accretion'. We suggest that the physical mechanism for this bursty accretion is a Hall-mediated instability of whistler waves in a differentially rotating fluid.  We also discuss the implications of the Hall effect for disc observables and the degree to which the Hall effect introduces a bi-modality in disc properties.

The paper is organized as follows. In Section~\ref{method}, we describe the equations of non-ideal MHD and the numerical algorithms we employ to solve them. We also detail our treatment of the ionization physics and chemistry and our choice of the simulation parameters used in our runs. Finally, we end this Section by defining the the relevant diagnostics computed from our simulations. In Section~\ref{results} we present our results and analysis, focusing on the inner-disc and outer-disc regions separately. Section~\ref{discussion} places our results within the context of current models for the structure and evolution of protoplanetary discs and discusses potential observational implications.  Our conclusions are summarized in Section~\ref{conclusions}.

\section{Method}
\label{method}

\subsection{Shearing-box equations}
\label{shearing_box}

Our numerical simulations are carried out in the local shearing-box approximation \citep{hawley95a}, a useful framework for describing phenomena that vary on lengthscales much less than the large-scale properties of the disc. 
A small patch of the disc, co-orbiting with angular velocity $\bm{\Omega} = \Omega_0 \ez$ at a fiducial radial location $R_0$, is represented in Cartesian coordinates $(x,y,z)$ with $x=R-R_0$ and $y=R_0\phi$ corresponding to the radial $(R)$ and azimuthal $(\phi)$ directions of a cylindrical coordinate system. Differential rotation is accounted for by including the Coriolis force and by imposing a background linear shear $\bm{v}_0 = -q\Omega_0 x\ey$, where
\[
q \equiv - \frac{1}{\Omega_0} \!\left. \deriv{\ln R}{\Omega(R)} \right|_{R=R_0} ;
\]
we use $q = 3/2$, appropriate for a Keplerian disc.

In units such that the magnetic permeability is equal to unity, the equations of motions are the continuity equation,
\begin{equation}
\label{continuity_eqn}
\D{t}{\rho} + \del \bcdot \bigl(\rho\bm{v}\bigr) = 0,
\end{equation}
the momentum equation,
\begin{eqnarray}
\label{momentum_eqn}
\lefteqn{
\D{t}{\rho\bm{v}} + \del \bcdot \bigl(\rho \bm{v}\bm{v} - \bm{B}\bm{B} \bigr) + \del \left( P + \frac{1}{2} B^2\right) 
}\nonumber\\*&&\mbox{}
= 2\rho q \Omega^2_0\bm{x} - \rho \Omega^2_0\bm{z} - 2 \Omega_0 \ez \btimes \rho \bm{v},
\end{eqnarray}
and the induction equation,
\begin{eqnarray}
\label{induction_eqn}
\lefteqn{
\D{t}{\bm{B}} - \del \btimes \bigl( \bm{v} \btimes \bm{B} \bigr)
}\nonumber\\*&&\mbox{}
= - \del \btimes \left[\etao \bm{J} + \etah \frac{\bm{J}\btimes\bm{B}}{B} - \etaa \frac{(\bm{J}\btimes\bm{B})\btimes\bm{B}}{B^2} \right],
\end{eqnarray} 
where $\rho$ is the mass density, $\rho \bm{v}$ is the momentum density, $\bm{B}$ is the magnetic field, $P$ is the gas pressure, and $\bm{J} = \del \btimes \bm{B}$ is the current density. We assume an isothermal equation of state, $P = \rho \cs^2$, where $\cs$ is the isothermal sound speed.  From left to right, the source terms in Equation~(\ref{momentum_eqn}) correspond to radial tidal forces (gravity and centrifugal), vertical gravity, and the Coriolis force. The electromotive force (EMF) on the right-hand side of Equation~(\ref{induction_eqn}) includes contributions from Ohmic diffusion (`O'), the Hall effect (`H'), and ambipolar diffusion (`A'). The importance of these non-ideal terms is characterized by the corresponding diffusivity parameters $\etao$, $\etah$, and $\etaa$, respectively, whose magnitudes we specify in Section \ref{parameters}.

\subsection{Numerical algorithm}
\label{numerics}

For our numerical calculations, we use a modified version of the PLUTO code \citep{mignone07a}.  PLUTO integrates the shearing-box equations (\ref{continuity_eqn})--(\ref{induction_eqn}) using a standard Godunov scheme with second-order--accurate spatial reconstruction, a monotonised central flux limiter, and a second-order--accurate Runge-Kutta time integration.  We use a modified HLL Riemann solver \citep[see Appendix A of][]{lesur14} to properly integrate the Hall effect in the induction equation (\ref{induction_eqn}). The remaining non-ideal MHD terms, Ohmic diffusion and ambipolar diffusion, are computed using second-order--accurate finite differencing and added to the induction equation in a manner consistent with the method of constrained transport \citep[CT;][]{evans88} to preserve the solenoidality constraint $\del \bcdot \bm{B} = 0$ to machine precision. Ohmic and ambipolar diffusion are treated via a slightly modified version of the resistivity algorithm of the public version of PLUTO to avoid repeated current calculations. The inclusion of the Hall effect requires slightly more care, as described in Appendix B of \citet{lesur14}. For all of the non-ideal effects, we use the same implementation as those authors.

The radial ($x$) boundary conditions are shearing periodic \citep{hawley95a}; the azimuthal boundaries are periodic; and the vertical boundary conditions satisfy a modified-outflow condition, in which $\rho$ is calculated based on hydrostatic equilibrium, $v_x$ and $v_y$ are extrapolated with zero vertical gradient, $v_z$ is a standard outflow condition, and ${\bmath B}$ is forced to be purely vertical. These boundary conditions have been shown to produce results similar to zero-current condition on the boundary \citep{lesur14}.

Shearing-box simulations of the MRI in vertically stratified discs generically exhibit significant mass outflow through the vertical boundaries of the computational domain \cite[e.g.][]{suzuki09,simon13b}. Anticipating this, and with a desire to achieve a steady-state solution in our simulations, we continually replenish the disc mass by multiplying the density in each cell at each time step by a constant factor chosen such that the total mass within the computational domain is held constant. While this globally breaks conservation of momentum in our simulations -- we do not decrease the velocity in each cell to compensate for the increased density -- any local patch of a realistic accretion disc will, almost by definition, have mass inflow from larger radii. Our intention in replenishing the mass lost to vertical outflows is to numerically mimic this physical situation in a computational domain whose horizontal boundary conditions are shearing-periodic and thus preclude radial mass inflow.

\subsection{Simulation parameters}
\label{parameters}

Most of our ionization profiles and simulation parameters are taken from \cite{lesur14}, and we refer the reader to that work for more details. Briefly, we adopt for our model disc the column-density and temperature profiles of the minimum mass solar nebula \citep[MMSN;][]{hayashi81}, which are given respectively by
\begin{equation}
\label{mmsn_surface_density}
\Sigma(R) = 1700 \left(\frac{R}{1~{\rm au}}\right)^{-3/2} {\rm g}~{\rm cm}^{-2},
\end{equation}
\begin{equation}
\label{mmsn_temperature}
T(R) = 280 \left(\frac{R}{1~{\rm au}}\right)^{-1/2} {\rm K} .
\end{equation}
The density at the mid-plane, $\rho(R)$, is determined by Equation (\ref{mmsn_surface_density}) and the relation $\Sigma = \sqrt{2\pi} \rho H$, where $H(R) \equiv \cs / \Omega$ is the thermal pressure scale height of the disc at radius $R$. For a disc consisting of 80\% H$_2$ and 20\% He by number orbiting a solar-mass protostellar object, this gives
\begin{equation}
\label{mmsn_mass_density}
\rho(R) = 1.4\times10^{-9} \left( \frac{R}{1~{\rm au}} \right)^{-11/4}~{\rm g~cm}^{-3} .
\end{equation}
In this paper, we present simulations carried out at a series of radii, ranging from $R_0 = 1~{\rm au}$ to  $R_0 = 100~{\rm au}$ and follow the dynamics up to a height $z = \pm 6H_0$, where $H_0$ is the disc scale height at $R_0$. Previously published simulations by some of the authors \citep{lesur14} focused on the evolution of the MRI at $R_0 = 1$, $5$, and $10~{\rm au}$. We include some of these simulation data alongside our own.

All of our model discs are initially in vertical hydrostatic equilibrium,
\begin{equation}
\label{initial_density}
\rho(x,y,z) = \rho(R_0) \, {\rm exp}\left(-\frac{z^2}{2H^2_0}\right) ,
\end{equation}
and are threaded by a uniform vertical magnetic field, $\bm{B} = B_0 \ez$, whose strength and direction is determined by specifying the dimensionless free parameter 
\begin{equation}
\beta_0 \equiv \frac{B_0}{|B_0|} \frac{ 2 \rho_0 c_{{\rm s}0}^2}{B^2_0} ,
\end{equation}
where $\rho_0$ is the initial density evaluated at the disc mid-plane. It is well established that the MRI, when subject to Hall EMFs, is sensitive to the relative orientation between the local magnetic field and the rotation axis \citep{wardle99a,balbus01,sano02a,sano02b}.\footnote{In fact, the Hall-MRI is also sensitive to the angle between the local magnetic field and the local vorticity \citep{kunz08}; for a Keplerian disc, the two of course coincide.} We thus explore both magnetic polarities, with $\beta_0 = \pm 10^3$, $\pm 10^4$, and $\pm 10^5$. The linear growth of the MRI is seeded by perturbing the radial velocity with a uniformly distributed random number between $\pm0.1\cs$.

We quantify the diffusivities via several dimensionless parameters that are independent of the magnetic-field strength (thus making them constant in time). For Ohmic resistivity, we define the magnetic Reynolds number
\begin{equation}
\rem \equiv \frac{\cs H}{\etao};
\end{equation}
ambipolar diffusion is quantified by its Elsasser number,
\begin{equation}
{\rm Am} \equiv \frac{\va^2}{\Omega_0 \etaa} .
\end{equation}
Following \cite{lesur14}, we define the Hall Reynolds number 
\begin{equation}
R_{\rm H} \equiv \frac{\va H}{\etah} \equiv \frac{H}{\ell_{\rm H}} ,
\end{equation}
where the Hall lengthscale $\ell_{\rm H}$ is defined {\it in situ}. (In an ion-electron-neutral plasma, it is equivalent to the ion skin depth divided by the square root of the mass-weighted ionization fraction).

%
%
\begin{figure}
\begin{center}
\leavevmode
\includegraphics[width=1.\columnwidth]{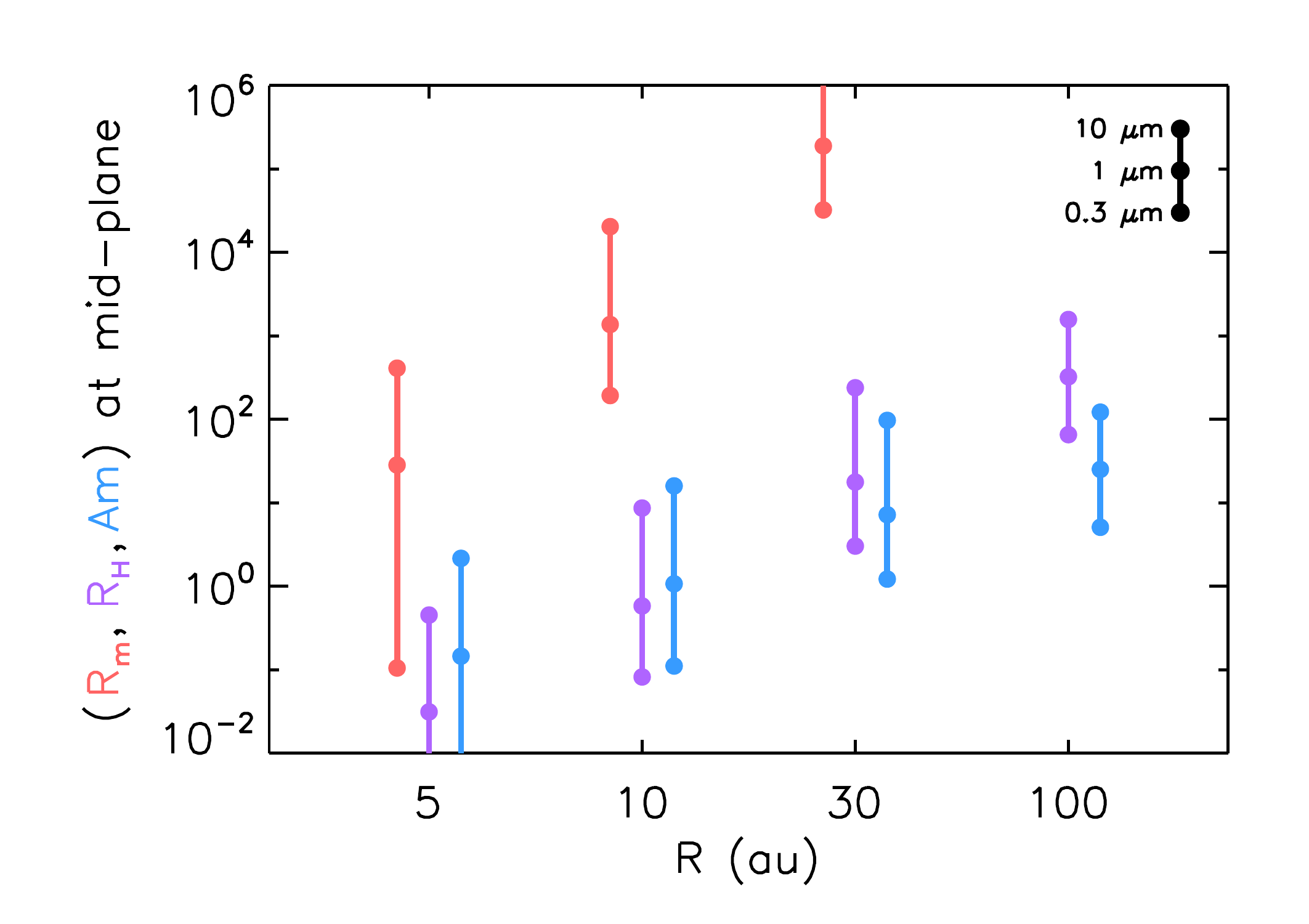}
\end{center}
\caption[]{Representative ranges of $\rem$, $R_{\rm H}$, and ${\rm Am}$ at different radii in the mid-plane of our model disc, which is ionized by cosmic rays at a rate $\zeta_{\rm cr} = 10^{-17}\exp(-\Sigma/96~{\rm g~cm}^{-2})$ \citep{umebayashi80} and which contains dust grains (1\% by mass) of varying radii ($0.3~\mu{\rm m}$, $1~\mu{\rm m}$, $10~\mu{\rm m}$). Other chemistry parameters (e.g.~grain material density, recombination rates, gas-grain collision rates) are taken from \citet{kunz09}.}
\label{fig:chemres}
\end{figure}

In choosing the profiles of the magnetic diffusivities $\etao$, $\etah$, and $\etaa$, we are guided by the chemical model and generalized Ohm's law prescribed in \S4 and appendix B of \citet{kunz09} and the setup used
in the shearing-box simulations of \cite{lesur14}.  In brief, for the simulations between $R_0 = 1$ and $10~{\rm au}$, we adopt the same ionization structure as \cite{lesur14}.  At larger radii, we choose mid-plane values of $\rem$, ${\rm Am}$, and $R_{\rm H}$ that are within the bounds calculated using the chemical model of \citet{kunz09} with varying dust grain size; see Table \ref{tbl:sims} and Fig.~\ref{fig:chemres}.  Note that in calculating our diffusivities, we have assumed that the mid-plane is ionized by cosmic rays $\zeta_{\rm cr} = 10^{-17}\exp(-\Sigma/96~{\rm g~cm}^{-2})$ \citep{umebayashi80}. Recent results have brought into question the viability of cosmic rays as an ionization source, as there are concerns about geometrical effects significantly attenuating cosmic rays \citep{umebayashi09} or the stellar wind blocking them altogether \citep{cleeves13,cleeves15}.  Thus, we emphasize that our setup is just one model of many and that the cosmic ray flux used here should be taken as an upper limit.

Away from the mid-plane, we adopt a simple approach to make contact with previous numerical setups \cite[e.g.][]{simon13a,simon13b} in which FUV photons are assumed to strongly ionize a thin layer above and below a diffusion-dominated mid-plane region. We set the diffusivities to be such that the corresponding Elsasser numbers (i.e.~$\va^2/\Omega_0 \eta$, where $\va$ is the $\alf$ speed and $\eta$ is the corresponding diffusivity) are initially constant (at their mid-plane values) for columns greater than $0.01~{\rm g~cm}^{-2}$ and assume a very high ionization fraction ($10^{-5}$) elsewhere \cite[following][]{perez-becker11b}. This ionization depth corresponds to $\sim 2 H_0$ away from the disk mid-plane at these radii. In a Courant-limited integration in which diffusion is important, the time step is inversely proportional to the diffusivity and thus computations can become prohibitively expensive. In these regions of high ionization, the Hall and Ohmic Elsasser numbers are $\gg 1$, but ambipolar diffusion, whose associated Elsasser number is proportional to gas density, can still be quite strong at large $|z/H|$. To minimize the effect of strong ambipolar diffusion on reducing the time step, we cap $\eta_{\rm A}$ in these regions to values $10\Omega_0 H^2_0$. \cite{lesur14} examined the effect of changing this cap on the simulations at 1~au and found no significant differences in their results;  thus, we do not anticipate that the exact value of this cap will strongly influence the results here. In the simulations at $100~{\rm au}$, we do not include Ohmic resistivity at all, as it is expected to be negligible in these regions.

%
%
\begin{table*}
\normalsize
\begin{center}
\caption{Shearing-box simulations\label{tbl:sims}}
\begin{tabular}{ @{}l|cccccccc}
\hline
Label&
Radius&
$\rem$&
$R_{\rm H}$&
${\rm Am}$&
$\beta_0$&
$\alpha$&
$\alpha_{\rm mid}$\\
 & (au) & & & & & & \\
\hline
\hline
1-OHA-5p-LKF		& $1$ 	& $2.9$ 		& $0.011$	& $0.24$	& $10^5$ 	& $5.0\times10^{-2}$ & $3.5\times10^{-2}$ \\
1-OHA-5n-LKF		& $1$ 	& $2.9$ 		& $0.011$	& $0.24$	& $-10^5$	& $3.9\times10^{-4}$ & $-9.0\times10^{-7}$ \\
1-OHA-3p-LKF		& $1$ 	& $2.9$		& $0.011$	& $0.24$	& $10^3$	& $3.1\times10^{-1}$ & $2.2\times10^{-1}$ \\
5-OHA-3n			& $5$ 	& $1400$		& $0.66$	& $2.5$	& $-10^3$	& $2.4\times10^{-2}$ & $2.2\times10^{-5}$ \\
5-OHA-4n			& $5$ 	& $1400$ 		& $0.66$	& $2.5$	& $-10^4$	& $5.0\times10^{-3}$ & $8.4\times10^{-4}$ \\
5-OHA-5p-LKF		& $5$	& $1400$ 		& $0.66$	& $2.5$	& $10^5$	& $2.0\times10^{-2}$ & $2.1\times10^{-2}$ \\
5-OHA-5n 		& $5$	& $1400$ 		& $0.66$	& $2.5$	& $-10^5$	& $3.3\times10^{-3}$ & $2.3\times10^{-3}$ \\
5-OHA-5n-diff 		& $5$ 	& $1400$ 		& $0.66$	& $0.25$	& $-10^5$	& $6.3\times10^{-4}$ & $3.1\times10^{-7}$ \\
5-OHA-5n-axi 		& $5$	& $1400$ 		& $0.66$	& $2.5$	& $-10^5$	& $9.3\times10^{-4}$ & $7.9\times10^{-7}$ \\
5-OA-5n 			& $5$	& $1400$ 		& $\infty$	& $2.5$	& $-10^5$	& $9.2\times10^{-4}$ & $1.0\times10^{-4}$ \\
10-OHA-5p-LKF	& $10$ 	& $10000$ 	& $1.8$	& $3.2$	& $10^5$	& $1.0\times10^{-2}$ & $9.3\times10^{-3}$ \\
10-OHA-5n 		& $10$ 	& $10000$ 	& $1.8$	& $3.2$	& $-10^5$	& $4.6\times10^{-3}$ & $3.5\times10^{-3}$ \\
30-OHA-3p 		& $30$ 	& $43000$ 	& $6.5$	& $1$	& $10^3$	& $4.0\times10^{-2}$ & $9.8\times10^{-3}$ \\
30-OHA-3n 		& $30$ 	& $43000$ 	& $6.5$	& $1$	& $-10^3$	& $2.5\times10^{-2}$ & $4.2\times10^{-4}$ \\
30-OHA-4p 		& $30$ 	& $43000$ 	& $6.5$	& $1$	& $10^4$	& $9.3\times10^{-3}$ & $1.2\times10^{-3}$ \\
30-OHA-4n 		& $30$ 	& $43000$ 	& $6.5$	& $1$	& $-10^4$	& $5.1\times10^{-3}$ & $1.2\times10^{-5}$ \\
30-OHA-5p 		& $30$ 	& $43000$ 	& $6.5$	& $1$	& $10^5$	& $2.5\times10^{-3}$ & $7.4\times10^{-4}$ \\
30-OHA-5n 		& $30$ 	& $43000$	& $6.5$	& $1$	& $-10^5$	& $1.3\times10^{-3}$ & $8.0\times10^{-5}$ \\
100-HA-4p-Am10 	& $100$ 	& $\infty$ 		& $65$	& $10$	& $10^4$	& $5.2\times10^{-2}$ & $2.3\times10^{-2}$ \\
100-HA-4n-Am10 	& $100$ 	& $\infty$ 		& $65$	& $10$	& $-10^4$	& $3.7\times10^{-2}$ & $1.3\times10^{-2}$ \\
100-HA-4p-Am1 	& $100$ 	& $\infty$ 		& $65$	& $1$	& $10^4$	& $1.3\times10^{-2}$ &$9.0\times10^{-4}$ \\
100-A-4p-Am1		& $100$ 	& $\infty$ 		& $\infty$	& $1$	& $10^4$	& $1.1\times10^{-2}$ &$4.7\times10^{-4}$ \\
\hline
\end{tabular}
\end{center}
\end{table*}

Our simulations are conducted in units such that $c_{{\rm s}0} = \Omega_0 = H_0 = \rho_0 = 1$. For $R_0 = 1$--$10~{\rm au}$, the computational domain is chosen to have size $L_x\times L_y\times L_z = 4\times8\times12$, resolved by $N_x\times N_y\times N_z = 64\times64\times192$ cells.  At radii larger than $10~{\rm au}$, the domain size is $L_x\times L_y\times L_z = 8\times16\times12$ and is resolved by $N_x\times N_y\times N_z = 128\times256\times192$ cells.  All of our simulations are listed in Table~\ref{tbl:sims}; following the convention of \cite{lesur14}, the runs are labelled by their radial location in our model disc, the non-ideal MHD terms that are included, and by the strength and orientation of the magnetic field. For example, the shearing box in run 30-OHA-4n is placed at a disc radius $R_0 = 30~{\rm au}$, includes the Ohmic, ambipolar, and Hall terms, and has $\beta_0 = -10^4$; the `n' in 4n indicates that $\bm{\Omega}\bcdot\bm{B} < 0$ initially (as opposed to `p' for $\bm{\Omega} \bcdot \bm{B} > 0$). Two of our simulations used a mid-plane value of ${\rm Am} = 1$ instead of ${\rm Am} = 10$ as calculated using the methods described above.  These runs are labelled with `Am1' appended to the simulation name.  Finally, the simulations taken directly from \cite{lesur14} are appended with `LKF'.

\subsection{Diagnostics}

We use several diagnostics to characterize the physics of accretion in our shearing-box simulations.  The first diagnostic is a volume average over the entire domain,
\begin{equation}
\label{volume_average}
\langle Q\rangle \equiv \frac{1}{L_xL_yL_z}\int\!\!\int\!\!\int {\rm d}x{\rm d}y{\rm d}z \, Q.
\end{equation}
We also perform horizontal averages to plot quantities in $(z,t)$ space-time diagrams; this average is denoted with angled brackets subscripted with $xy$:
\begin{equation}
\label{horizontal_average}
\langle Q\rangle_{xy} \equiv \frac{1}{L_xL_y}\int\!\!\int {\rm d}x {\rm d}y \, Q.
\end{equation}
Finally, we denote a time average with an over-bar,
\begin{equation}
\label{time_average}
\overline{Q} = \frac{1}{\tau}\int {\rm d}t\, Q ,
\end{equation}
where $\tau$ is the time range over which we average. For each simulation, this averaging is chosen to begin after initial transients have died out and to end when the simulation is terminated.

Throughout much of this work, we examine the density-weighted $R\phi$-component of the stress tensor, $ \rho v_x \delta v_y - B_xB_y$, which is responsible for the radial transport of angular momentum. We cast this quantity in terms of the $\alpha$ parameter of \cite{shakura73} by dividing by the gas pressure and performing the appropriate averages,
\begin{equation}
\label{alpha}
\alpha \equiv \overline{\left[\frac{\left\langle \rho v_x \delta v_y
- B_xB_y\right\rangle}{\left\langle \rho\cs^2\right\rangle}\right]}.
\end{equation}
The value of $\alpha$ for each simulation is given in Table~(\ref{tbl:sims}), along with the same quantity integrated only over the disc mid-plane ($|z| \le 0.5$), $\alpha_{\rm mid}$.

While angular momentum can also be transported in the vertical direction by wind-launching mechanisms \cite[e.g.][]{blandford82,bai13a,bai13b}, this angular-momentum transport is not well defined in a shearing box. As pointed out by several  authors \cite[e.g.][]{simon13b,bai13b,lesur13}, the radial symmetry of the shearing box can lead to magnetic-field structures that launch outflows from the top and bottom of the box in opposite radial directions. Furthermore, the fast magnetosonic point is not located within local domains \citep{lesur13}. Therefore, we will not discuss this angular-momentum transport mechanism further.

\section{Results}
\label{results}

\subsection{Inner disc}
\label{inner_disc}

\subsubsection{Simulations results}
\label{inner_sims}

\begin{figure}
\begin{center}
\leavevmode
\includegraphics[width=1.\columnwidth]{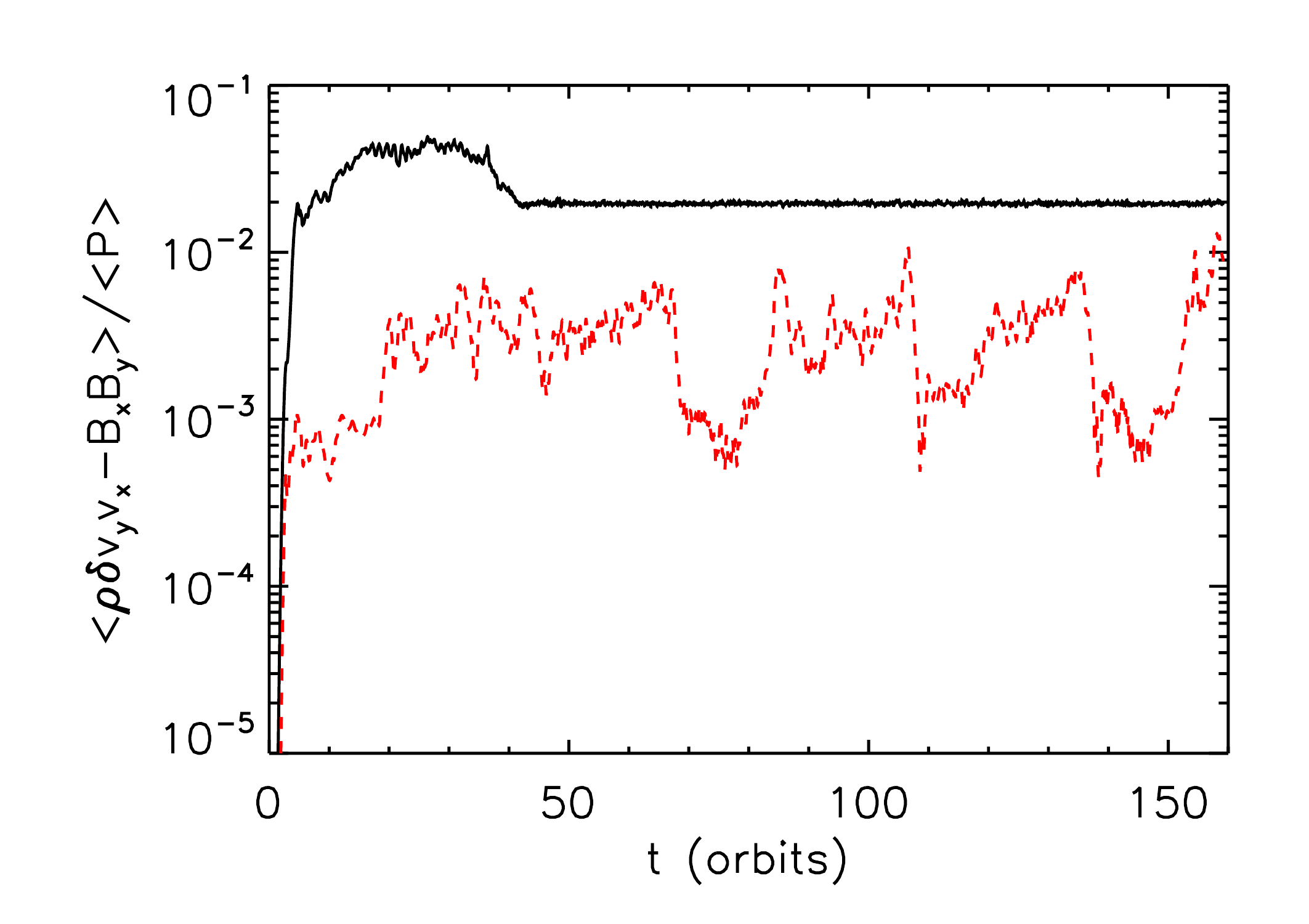}
\end{center}
\caption[]{Volume-averaged total (Maxwell plus Reynolds) stress normalized by the volume-averaged gas pressure versus time. The black, solid line is run 5-OHA-5p ($5~{\rm au}$, $\beta_0 = 10^5$) and the red, dashed line is 5-OHA-5n ($5~{\rm au}$, $\beta_0 = -10^5$). The case with negative $\beta_0$ shows significant temporal variations of up to an order of magnitude in amplitude, whereas the positive $\beta_0$ run shows very flat accretion stress after an initial transient phase.}
\label{stress_r5}
\end{figure}

In this Section, we consider radii $1$--$10~{\rm au}$, within which the Hall effect is the dominant non-ideal process \citep{wardle99b,sano02a,wardle07}. \cite{lesur14} showed that if $\bm{\Omega}\bcdot\bm{B} > 0$, the typical $\alpha$ values are in the range $0.01$--$0.1$. Furthermore, the Maxwell stress is dominated by a laminar structure that exists at the largest horizontal scales of the box rather than by fluctuations that exist on smaller scales. \cite{lesur14} also considered the case $\bm{\Omega}\bcdot\bm{B} < 0$ for $R_0 = 1~{\rm au}$ (which we have included in our simulation suite) and found essentially no transport.

\begin{figure*}
\begin{center}
\leavevmode
\includegraphics[width=1.9\columnwidth]{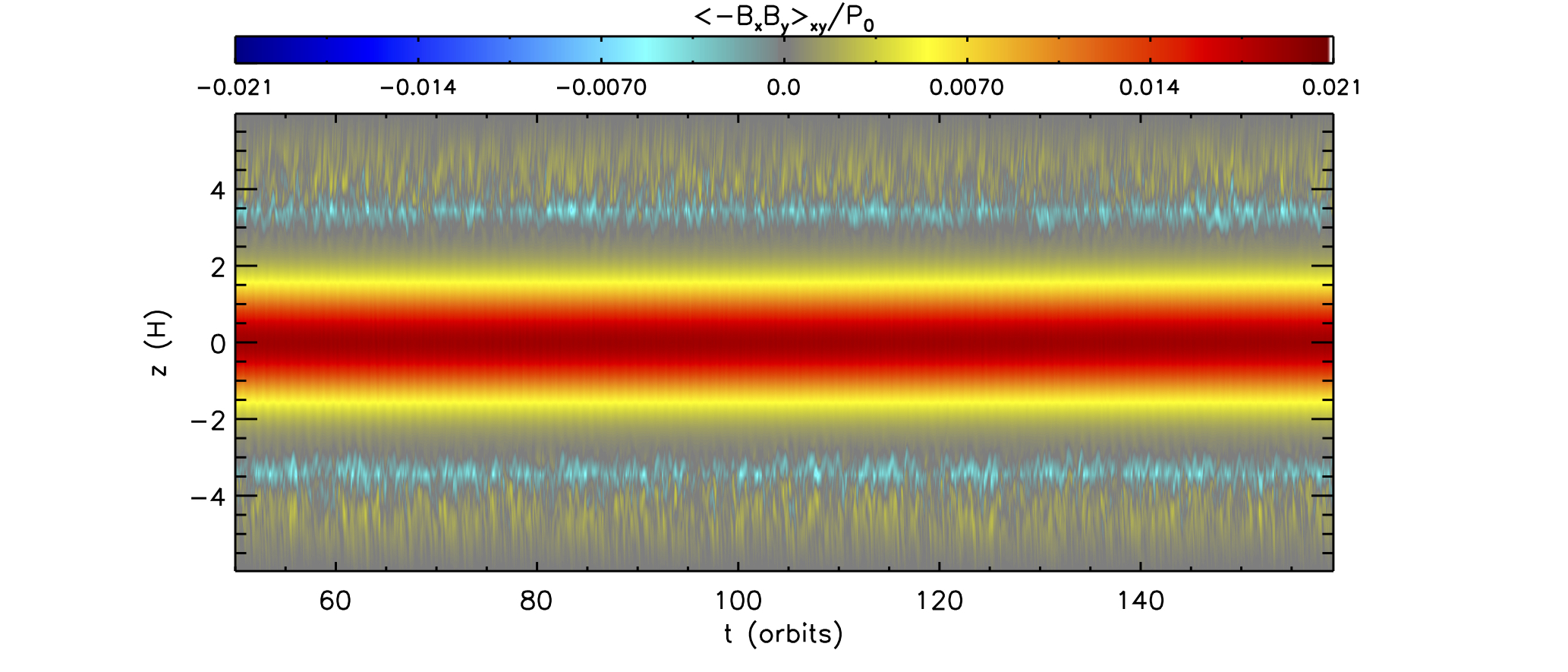}
\includegraphics[width=1.9\columnwidth]{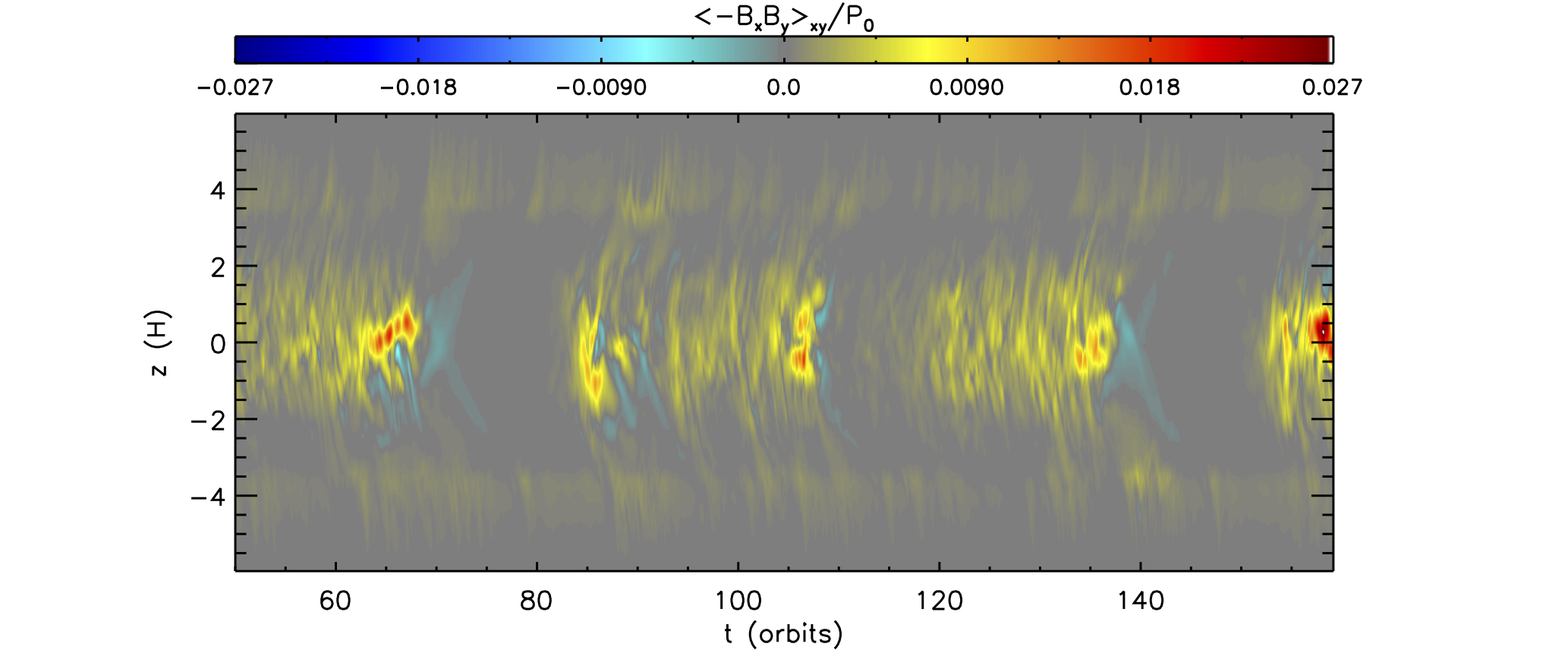}
\end{center}
\caption[]{Space-time diagram of the horizontally averaged Maxwell stress, normalized by the initial, mid-plane gas pressure for the two runs shown in Fig.~\ref{stress_r5}.  The top panel corresponds to 5-OHA-5p ($5~{\rm au}$, $\beta_0 = 10^5$) and the bottom panel corresponds to 5-OHA-5n ($5~{\rm au}$, $\beta_0 = -10^5$). The $\beta_0 < 0$ case shows significant temporal variations in the Maxwell stress that occur near the disc mid-plane. The $\beta_0 > 0$ run shows a constant Maxwell stress in this same region.} 
\label{sttz_max_r5}
\end{figure*}

We have expanded on these inner-disc simulations by carrying out simulations at $5$ and $10~{\rm au}$ with $\bm{\Omega}\bcdot\bm{B} < 0$. The evolution of the volume-averaged stress is shown in Fig.~\ref{stress_r5} for two simulations at $5~{\rm au}$ with $\beta_0 = 10^5$ (black, solid line) and $\beta_0 = -10^5$ (red, dashed line). In contrast to the stress measured in the $\bm{\Omega}\bcdot\bm{B} < 0$ simulations at $1~{\rm au}$, the stress measured at $5~{\rm au}$ undergoes significant temporal fluctuations for $\bm{\Omega}\bcdot\bm{B} < 0$, varying by up to an order of magnitude. Fig.~\ref{sttz_max_r5} shows the space-time evolution of the Maxwell stress for these same two runs. In both cases, the Maxwell stress is dominant near the disc mid-plane, but for $\bm{\Omega}\bcdot\bm{B} > 0$ the stress is almost completely dominated by the largest horizontal scales in the box (i.e., $k_x = k_y = 0$), whereas the bursts of stress in the $\bm{\Omega}\bcdot\bm{B} < 0$ case have a non-negligible small-scale component. Given analytic expectations that an anti-aligned magnetic field in an accretion disc subject to strong Hall diffusion should be MRI-stable \citep[e.g.][]{wardle99a}, such bursty turbulence with $\alpha_{\rm mid}$ ranging from $\sim$$10^{-3}$ up to $\sim$$10^{-2}$ is rather surprising. Evidently, {\em even for} $\bm{\Omega}\bcdot\bm{B} < 0$ {\em in the Hall-dominated regime, it is possible to get significant amounts of turbulent activity and angular-momentum transport near the disc mid-plane.}

\begin{figure*}
\centering
\leavevmode
\includegraphics[width=1.9\columnwidth]{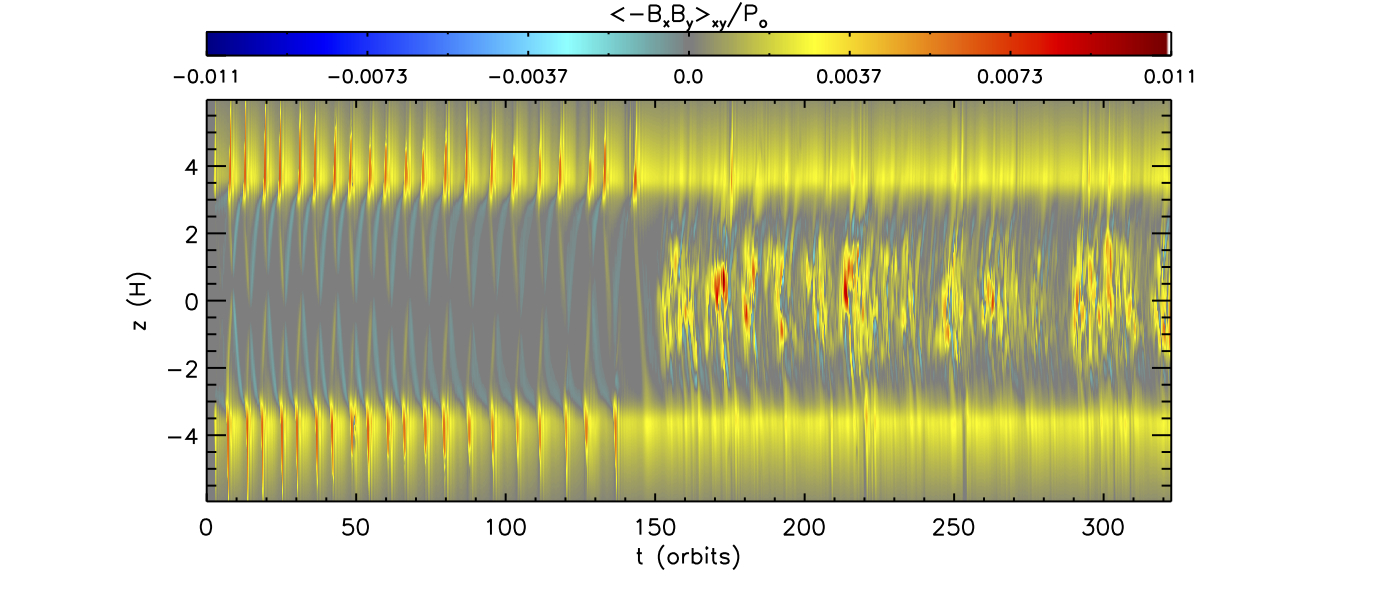}
\caption[]{Space-time diagram of the horizontally averaged Maxwell stress, normalized by the initial, mid-plane gas pressure for the run 5-OHA-4n.  There appear to be two separate regimes of evolution.  Until $\sim$$150$ orbits, the Maxwell stress near the disc mid-plane is small, but in the upper atmosphere varies substantially.  After $150$ orbits, the variability in the active regions is reduced and there appear to be spasmodic increases in Maxwell stress within the mid-plane region.}
\label{sttz_max_r4n}
\end{figure*}

To better understand the conditions under which such variable accretion can occur, we performed several additional inner-disc simulations with $\bm{\Omega}\bcdot\bm{B} < 0$. We found that this variability is also seen for $\beta_0 = -10^5$ at $10~{\rm au}$, suggesting that the exact values of the diffusion coefficients do not matter so long as the Hall effect is the dominant term in the magnetic induction equation and ${\rm Am}$ is larger than unity. We also checked the dependence of this bursty behaviour on the strength of the magnetic field.  The equivalent $5~{\rm au}$ run with $\beta_0 = -10^4$ (5-OHA-4n) shows similar variability, though the magnitude of fluctuations is less than an order of magnitude. Fig.~\ref{sttz_max_r4n} shows the space-time evolution of the Maxwell stress for this run.  The simulation spends a significant time in a state with very low stress near the mid-plane and highly variable stress in the upper atmosphere, followed by a more temporally fluctuating stress in the mid-plane starting after $\sim$150 orbits. The simulation with $\beta_0 = -10^3$ at $5~{\rm au}$ (run 5-OHA-3n) shows no Maxwell stress within the mid-plane region and significant large-scale stress for $|z| \gtrsim 2$.

Finally, we recall that this large temporal variability was not seen in previous numerical studies of the inner-disc region. For example, \citet{lesur14} carried out $\bm{\Omega}\bcdot\bm{B} < 0$ simulations at $1~{\rm au}$ and found the angular-momentum transport to be significantly reduced ($\alpha \sim 10^{-4}$ and no significant outflow). In their disc model at $1~{\rm au}$, Ohmic dissipation is as strong as the Hall effect, suggesting that the Hall effect must be dominant to trigger such bursty accretion. \citet{bai14a} explored $\bm{\Omega}\bcdot\bm{B} < 0$ at $5~{\rm au}$, where the Hall effect does dominate over Ohmic dissipation, but used quasi-1D shearing boxes with negligible horizontal dimensions. This suggests that the turbulent activity we see in such a case requires 
either $k_x \ne 0$, $k_y \ne 0$ or both.  We test the conjecture that the burst activity requires non-axisymmetry ($k_y \ne 0$) by carrying out a test simulation in which axisymmetry was enforced by setting $N_y = 1$ (run 5-OHA-5n-axi). This run showed no sign of bursty behaviour and the mid-plane region remained relatively quiescent. It is also worth noting that run 5-OHA-3n, which does not exhibit such highly variable stress (for reasons that remain unclear), also does not exhibit significant non-axisymmetric fluctuations. Follow-up work by \citet{bai15} did explore the case $\bm{\Omega}\bcdot\bm{B} < 0$ at both $5$ and $10~{\rm au}$ in full non-axisymmetric 3D, but did not find the variability we see in our corresponding simulations. 
While \citet{bai15} used a different chemistry calculation than the one employed here, resulting in the enhancement of all three non-ideal effects, we suspect that the absence of bursts in that work is specifically due to the enhancement of ambipolar diffusion, resulting in a relatively smaller value of ${\rm Am}$ compared to our work. To test this hypothesis, we carried out an identical simulation to 5-OHA-5n but with the magnitude of $\etaa$ amplified by a factor of $10$ everywhere; this run is labelled 5-OHA-5n-diff in Table~\ref{tbl:sims}. We found no turbulent activity in the mid-plane for this test simulation. While it is unsurprising that a decrease in ${\rm Am}$ leads to less activity in the disc, it is important to point out that the value of ${\rm Am}$ that we have used in our fiducial simulations falls within the range of diffusion values based on our chemistry calculations (see Fig.~\ref{fig:chemres}).

To summarize, we find that inner-disc regions of protoplanetary discs in which a net vertical magnetic field is anti-aligned with the rotation axis, the Hall effect dominates over magnetic induction and the other non-ideal processes, and neutrals collide at least a few times per orbit with free charges can support appreciable amounts of highly variable turbulent transport. What makes this conclusion all the more startling is that we find such disc regions to exhibit larger $\alpha$ than they would if the Hall effect were otherwise absent (see run 5-OA-5n in Table \ref{tbl:sims})! To aid in our interpretation of this behaviour, we return to the linear theory.

\subsubsection{Origin of the turbulent bursts}
\label{inner_origin}

To provide some illumination on the observed bursty behaviour seen in some of our simulations, we revisit the general linear analysis of the Hall effect in protoplanetary discs \citep[cf.][]{balbus01}. We begin by perturbing Equations (\ref{continuity_eqn})--(\ref{induction_eqn}) about the equilibrium state $\rho = \rho_0$, $\bm{v} = -q \Omega_0 x \ey$, and $\bm{B} = B_{y0} \ey + B_{z0} \ez$, the latter taken to be constant. Eulerian perturbations are denoted by a $\delta$ and are taken to have space-time dependence $\propto$$\delta(t) \exp( \imag \bm{k}\bcdot\bm{x} )$, where the time-dependent wavevector $\bm{k} = \bm{k}(t)$ satisfies 
\begin{equation}
\label{eq:kdot}
\deriv{t}{\bm{k}} = q\Omega_0 k_y \ex .
\end{equation}
To keep the analysis relatively simple, we shall ignore Ohmic dissipation and ambipolar diffusion and work in the incompressible limit at the disc mid-plane. The latter assumption precludes sound waves and (stable) buoyant oscillations. For notational convenience, we set $4\pi\rho_0 = 1$, so that the Alfv\'{e}n velocity in the equilibrium state $\bm{v}_{{\rm A},0} = \bm{B}_0$. 

Under these assumptions, Equations (\ref{continuity_eqn})--(\ref{induction_eqn}) written out to linear order in $\delta$ are, respectively,
\begin{align}
\label{eq:conlin}
0 &= \bm{k}\bcdot\delta\bm{v} , \\
\label{eq:momlin}
\deriv{t}{\delta\bm{v}} &=-i\bm{k} \,\delta\Pi + 2 \Omega_0 \delta v_y \ex - (2-q) \Omega_0 \delta v_x \ey + \imag\oA\delta\bm{B} , \\
\label{eq:indlin}
\deriv{t}{\delta\bm{B}} &= - q\Omega_0 \delta B_x\ey + \imag \oA \delta\bm{v}  + \ell_{\rm H} \oA \bm{k}\btimes\delta\bm{B} ,
\end{align}
where we have introduced the compact notation $\oA \equiv \bm{k}\bcdot\bm{v}_{{\rm A}0}$ for the Alfv\'{e}n frequency and $\delta\Pi \equiv \delta P + \bm{B}_0 \bcdot\delta\bm{B}$ for the perturbed total (gas + magnetic) pressure. Since ${\rm d}k_x/{\rm d}t = q\Omega_0 k_y$ (Eq.~\ref{eq:kdot}), Equation (\ref{eq:indlin}) together with Equation (\ref{eq:conlin}) guarantee the divergence-free condition ${\rm d}(\bm{k}\bcdot\delta\bm{B})/{\rm d}t = 0$. Note that the combination $\bm{k}\bcdot\bm{B}_0$ is also time-independent.

Motivated by the results described in Section \ref{inner_sims}, we consider non-axisymmetric fluctuations, $k_y \ne 0$. To make analytical progress, we assume $k_x$ evolves slowly compared to the shear timescale $(q\Omega_0)^{-1}$, which requires $k_y / k_x \ll 1$. The coefficients in Equations (\ref{eq:momlin}) and (\ref{eq:indlin}) are then constant to leading order and we may consider solutions of the form $\delta(t) = \delta \exp(\gamma t)$. Using Equation (\ref{eq:conlin}) in the $z$-component of Equation (\ref{eq:momlin}) to determine the perturbed pressure, the $x$- and $y$-components of the momentum equation (\ref{eq:momlin}) can be then re-written in the following compact matrix form:
\begin{equation}\label{eq:matrixmom}
\begin{pmatrix}
\delta v_x \\
\delta v_y 
\end{pmatrix}
=\frac{\imag\oA}{\gamma^2+\kappa^2\chi}
\begin{pmatrix}
\gamma & 2\Omega_0\chi \\
-(2-q)\Omega_0 & \gamma
\end{pmatrix}
\begin{pmatrix}
\delta B_x \\
\delta B_y 
\end{pmatrix} ,
\end{equation}
where $\kappa^2 \equiv 2 ( 2 - q ) \Omega^2_0$ is the square of the epicyclic frequency and $\chi \equiv k^2_z/k^2$ is a geometrical factor accounting for the anisotropy of the inertial response. Using Equation (\ref{eq:matrixmom}) to replace the perturbed velocity in the induction equation (\ref{eq:indlin}), we obtain
\begin{eqnarray}\label{eq:matrixind}
\lefteqn{
\left[
\begin{pmatrix}
\gamma & k_z v_{\rm H} \\
q\Omega_0 - k_z v_{\rm H} \chi^{-1} & \gamma
\end{pmatrix}
\right.
}\nonumber\\*&&\mbox{}
\left. + \frac{\oA^2}{\gamma^2+\kappa^2\chi}
\begin{pmatrix}
\gamma & 2\Omega_0 \chi \\
-(2-q)\Omega_0 & \gamma
\end{pmatrix} 
\right]
\begin{pmatrix}
\delta B_x \\
\delta B_y 
\end{pmatrix} = 0,
\end{eqnarray}
where $v_{\rm H} \equiv \oA \ell_{\rm H}$ is a scale-dependent Hall velocity. 

Setting the determinant of the $2\times2$ matrix in equation (\ref{eq:matrixind}) to zero results in the dispersion relation, which may be conveniently written as
\begin{eqnarray}\label{eq:disprel}
\lefteqn{
\left( \gamma^2 + \imag \gamma k v_{\rm H} + \oA^2 \right) \!\left( \gamma^2 - \imag \gamma k v_{\rm H} + \oA^2 \right) - k_z v_{\rm H} q \Omega_0 \left( \gamma^2 + \oA^2 \right)
}\nonumber\\*&&
\mbox{} = - \kappa^2 \chi \widetilde{\gamma}^2 + 4 \Omega_0 \chi \oA^2 \left( \Omega_0 - \frac{k_z v_{\rm H}}{\chi} \right) ,
\end{eqnarray}
where
\begin{equation}
\widetilde{\gamma}^2 \equiv \gamma^2 + k^2 v^2_{\rm H} + \oA^2 - k_z v_{\rm H} q \Omega_0.
\end{equation}
The corresponding eigenvectors are 
\begin{subequations}\label{eigenmode}
\begin{align}
\delta v_y &= - \frac{\xi_x}{\widetilde{\gamma}^2} \left[ (2-q) \Omega_0 \widetilde{\gamma}^2 - \oA^2 \left( 2\Omega_0 - \frac{k_z v_{\rm H}}{\chi} \right) \right] ,\\
\frac{\delta B_x}{\oA} &= + \frac{\imag\xi_x}{\widetilde{\gamma}^2} \left[ \widetilde{\gamma}^2 + k_z v_{\rm H} \left( 2\Omega_0 - \frac{k_z v_{\rm H}}{\chi} \right) \right]  ,\\
\frac{\delta B_y}{\oA} &= - \frac{{\rm i}\xi_x}{\widetilde{\gamma}^2} \left[ \gamma \left( 2\Omega_0 - \frac{k_z v_{\rm H}}{\chi} \right) \right],
\end{align}
\end{subequations}
where $\xi_x \equiv \delta v_x / \gamma$ is the radial displacement of a fluid element. We examine different limiting cases of the dispersion relation (\ref{eq:disprel}) and the eigenvectors (eq.~\ref{eigenmode}), not only to make contact with the existing literature but also to elucidate the various physical processes at play.

Let us first take the limit $\Omega_0\rightarrow 0$ (i.e.~no rotation, no shear), in which case Equation (\ref{eq:disprel}) becomes
\begin{equation}
\gamma^2 \mp \imag \gamma k v_{\rm H} + \oA^2 = 0 .
\end{equation}
The positive-frequency solutions ($\omega \equiv -\imag \gamma > 0$) are given by
\begin{align}
\label{eq:whistlers}
\frac{\omega}{\oA} = \mp \,\frac{k\ell_{\rm H}}{2} + \sqrt{1 + \left( \frac{k\ell_{\rm H}}{2}\right)^2} .
\end{align}
For small wavenumbers ($k\ell_{\rm H} \ll 1$), both of the above solutions reduce to Alfv\'{e}n waves; at large wavenumbers ($k\ell_{\rm H} \gg 1$), right-handed waves (plus sign) go over to the high-frequency whistler-wave branch ($\omega \approx k v_{\rm H}$), whereas large-$k$ left-handed ion-cyclotron waves (minus sign) are cut off at a frequency
\begin{equation}\label{eq:oH}
\omega_{\rm H} \equiv \frac{\oA^2}{kv_{\rm H}} = \frac{v_{{\rm A},0}}{\ell_{\rm H}} \left| \frac{\bm{k}\bcdot\bm{B}_0}{k B_0} \right| .
\end{equation}
With solid-body rotation included, the dispersion relation (\ref{eq:disprel}) reduces neatly to
\begin{equation}
\gamma^2 \mp \imag \gamma k v_{\rm H} \left( 1 - \frac{2\Omega_0\chi}{k_z v_{\rm H}} \right) + \oA^2 + 2\Omega_0 k_z v_{\rm H} = 0 ,
\end{equation}
and the solutions (\ref{eq:whistlers}) become
\begin{equation}\label{eq:rotwhistlers}
\frac{\omega}{\oA} = \mp \frac{k\widetilde{\ell}_{\rm H}}{2} + \sqrt{ 1 + \biggl( \frac{k\widetilde{\ell}_{\rm H}}{2} \biggr)^2  + \frac{2\Omega_0}{\omega_{\rm H}}\frac{k_z}{k} } ,
\end{equation}
where $\widetilde{\ell}_{\rm H} \equiv \ell_{\rm H} ( 1 - 2\Omega_0 \chi / k_z v_{\rm H} )$ is the Hall lengthscale modified by rotation -- its presence signals the interplay between the dynamical epicycles in the bulk neutral fluid and the `magnetic epicycles' (i.e.~circular polarization) induced by the Hall effect. Depending upon the orientation of the mean magnetic field, this circular polarization can increase or decrease the effective Coriolis force; e.g.~for $\bm{\Omega}\bcdot\bm{B} > 0$, the dynamical epicycle is slowed by Hall currents. In this case, the return magnetic tension force is effectively increased, a consequence of the last term in the square root in Equation (\ref{eq:rotwhistlers}). (See \S 3 of \citealt{balbus01} for an alternate discussion of this case.)

Next, we restore the differential rotation and take the ideal-MHD limit ($v_{\rm H}\rightarrow 0$). Equation (\ref{eq:disprel}) then reverts to the standard MRI dispersion relation of \citet{balbus91},
\begin{equation}
\left( \gamma^2 + \oA^2 \right)^2 = - \kappa^2\chi \left( \gamma^2 + \oA^2 \right) + 4\Omega^2_0 \chi \oA^2 ,
\end{equation}
for which unstable solutions exist if $\oA^2 - 2q \Omega^2_0 \chi < 0$. Non-axisymmetric shearing waves are amplified strongest by the MRI when $\chi \sim 1$. As these waves sweep from leading to strongly trailing, $\chi \rightarrow 0$ and the final (destabilizing) term becomes smaller and smaller. Eventually, the non-axisymmetric MRI is stabilized when $\chi < \oA^2 / 2q\Omega^2_0$. Hence, the ideal non-axisymmetric MRI is a transient instability \citep{balbus92}.

Finally, including both the Hall effect and differential rotation leads to two additional instabilities, which are physically distinct from the MRI. The first is the Hall-shear instability (HSI), which may be readily obtained by letting $(\oA/\Omega_0)^2 \sim R_{\rm H}/\sqrt{\beta} \ll 1$ in Equation (\ref{eq:matrixind}), thereby severing the dynamical link between the induction and momentum equations \citep{rudiger05,pandey12}. The dispersion relation (\ref{eq:disprel}) then reads
\begin{equation}
\left( \gamma^2 + \kappa^2 \chi \right) \left( \gamma^2 + k^2 v^2_{\rm H} - k_z v_{\rm H} q \Omega_0 \right) = 0.
\end{equation}
The first factor in parentheses represents epicyclic motion in the bulk neutral fluid, an inherently stabilizing influence in a Keplerian disc. The second factor in parentheses represents whistler waves, which are physically decoupled from the epicycles and may be driven unstable if $| q\Omega_0 / \omega_{\rm H} | > 1$, i.e.~if the time for an Alfv\'{e}n wave to travel a distance $\ell_{\rm H}$ is longer than the time for a magnetic perturbation to grow by the Keplerian shear. In this case, Equation (\ref{eigenmode}) gives
\begin{equation}\label{hsimode}
\frac{\delta B_x}{\delta B_y} = - \left( \frac{q\Omega_0}{k_z v_{\rm H}} - \frac{1}{\chi} \right)^{-1/2}  \sim \mathcal{O}(1) .
\end{equation}
This instability, and its non-axisymmetric guise, was also found by \citet{kunz08} when specializing the linear analysis to planar (i.e.~non-rotating) shear flows. It is important to note that the stabilizing term, $k^2 v^2_{\rm H}$, for a shearing wave eventually dominates over the destabilizing term, $k_z v_{\rm H} q \Omega_0$. Hence, the non-axisymmetric HSI is also a transient instability.

The second instability may be extracted by setting $k_z/k_x\equiv \ep\ll 1$, so that $\chi \sim \ep^2$. This asymptotic ordering precludes both the MRI and the HSI from entering the analysis, as it puts the focus on strongly leading/trailing disturbances. Employing the {\em Ansatz} $\gamma \sim \mathcal{O}(\ep)$, we find that the dispersion relation (\ref{eq:disprel}) at $\mathcal{O}(1)$ may be factored as
\begin{equation}
\gamma^2 k^2 v^2_{\rm H} + \left[ \oA^2 + 2\Omega_0 k_z v_{\rm H} \right] \! \left[ \oA^2 + (2-q)\Omega_0 k_z v_{\rm H} \right] = 0 .
\end{equation}
For $q > 0$, the perturbations are unstable when
\begin{equation}
\label{condition}
-\frac{\oA^2}{(2-q)\Omega_0 }< k_z v_{\rm H} < - \frac{\oA^2}{2\Omega_0},
\end{equation}
which, if satisfied for one $k_x$, is always satisfied. For this reason, non-axisymmetric shearing waves are linearly unstable without bound.\footnote{Recall that we have ignored the effects of Ohmic dissipation and ambipolar diffusion in our linear analysis, which would attenuate the growth of strongly leading/trailing shearing waves.} In this case, Equation (\ref{eigenmode}) gives
\begin{equation}\label{dmrimode}
\frac{\delta B_x}{\delta B_y} = - \frac{\kappa\sqrt{\chi}}{2\Omega_0} \, \left| { {\displaystyle k_z v_{\rm H} + \frac{\oA^2}{(2-q)\Omega_0} } \over {\displaystyle k_z v_{\rm H} + \frac{\oA^2}{2\Omega_0} } } \right|^{1/2} \sim \mathcal{O}(\epsilon) .
\end{equation}
In the limit $\Omega_0 \rightarrow 0$, this dispersion relation recovers the ion-cyclotron wave, $\gamma^2 \simeq -\oA^4 / k^2 v^2_{\rm H}$ (see Eq.~\ref{eq:oH}); thus, the instability arises from the interaction between ion-cyclotron waves and epicyclic motion in a differentially rotating disc. The axisymmetric version of this particular instability is referred to by \citet{pandey12} as the `diffusive MRI', because the growing azimuthal field is generated not by the differential shear of the radial field (as for the standard MRI and the HSI) but rather by Hall currents.

\subsubsection{Which instability might be responsible for the observed turbulent bursts?}\label{instability}

\begin{figure*}
\centering
\leavevmode
\includegraphics[width=\columnwidth,angle=90]{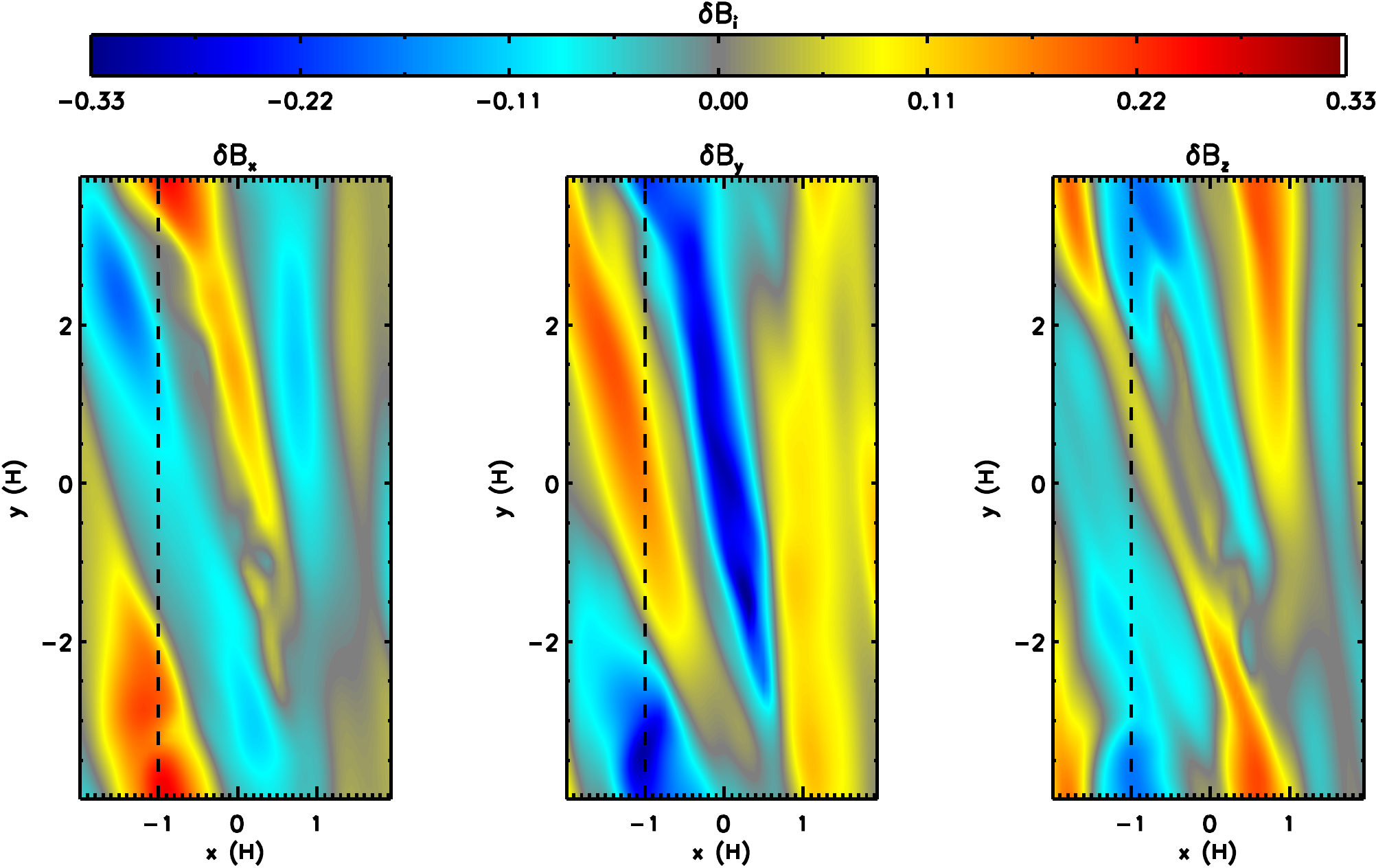}
\includegraphics[width=0.72\columnwidth,angle=90]{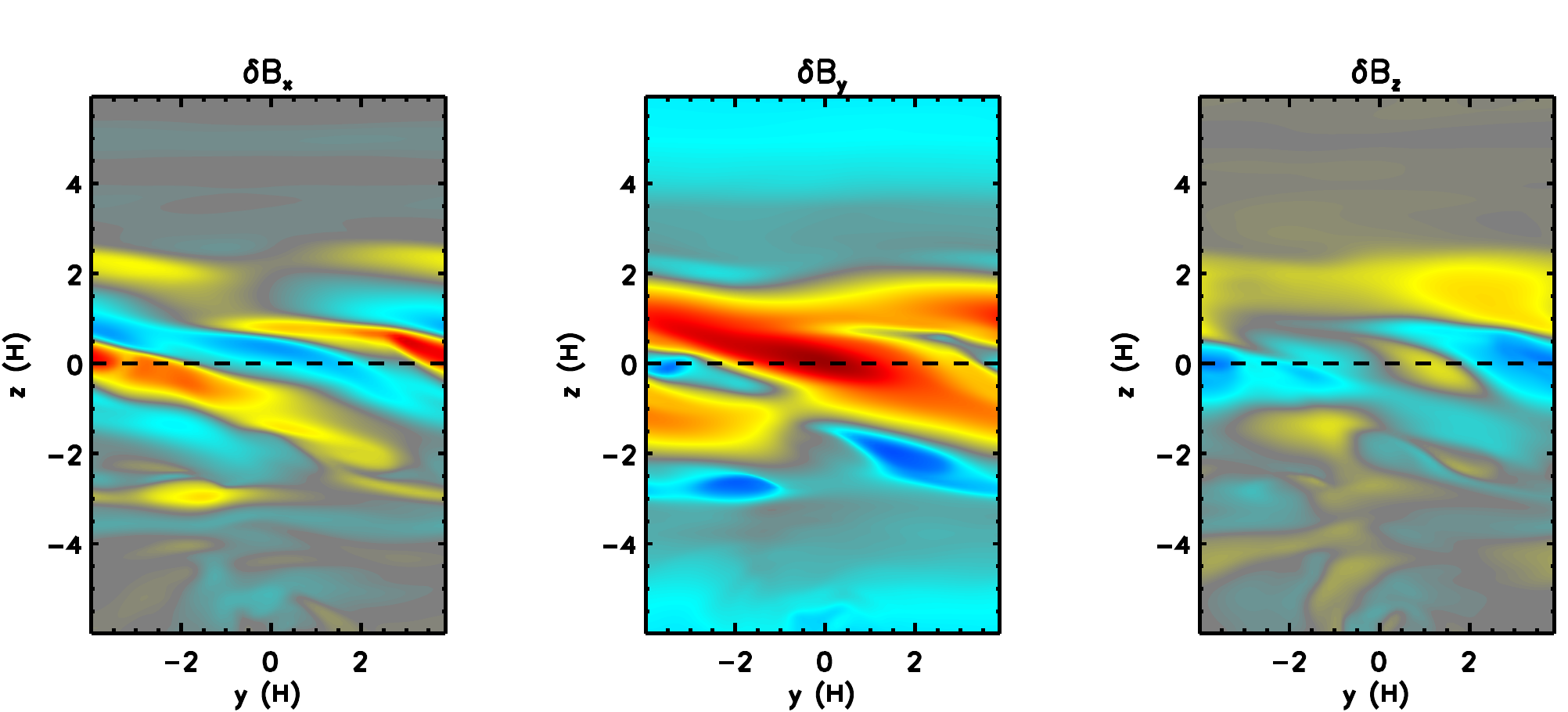}
\caption[]{Magnetic-field fluctuations in ({\it top row}) the $x$-$y$ plane at $z = 0$ and ({\it bottom row}) the $y$-$z$ plane at $x=-1$ at orbit number 157 in run 5-OHA-5n, during the largest-amplitude burst of Maxwell stress. From left to right, the panels show the $x$-, $y$-, and $z$-components of the fluctuating magnetic field $\delta\bm{B} = \bm{B} - \langle \bm{B}(x,y,z=0) \rangle_{xy}$; $\langle B_x \rangle_{xy} = -0.021$, $\langle B_y \rangle_{xy} = 0.26$, and $\langle B_z \rangle_{xy} = -0.0045$. The dashed line in the top (bottom) row indicates the planar cut shown in the bottom (top) row. The fluctuations exhibit non-axisymmetry with $k_z v_{\rm H} q \Omega_0 > 0$ and $(\oA/\Omega_0)^2 \ll 1$, implicating the non-axisymmetric HSI as the source of stress (\S\ref{instability}). These characteristics are generic to all of the bursts that we observed in this run.}
\label{bxy_r5}
\end{figure*}

The linear theory elucidated, we now return to those simulations exhibiting bursts of turbulent activity and analyze them with the aim of determining whether any of the instabilities described in the preceding section might be associated with the bursts. Our starting point is Fig.~\ref{bxy_r5}, which provides pseudo-color plots of the $x$-, $y$-, and $z$-components of the magnetic-field fluctuations $\delta \bm{B} = \bm{B} - \langle \bm{B} \rangle_{xy}$ (where the horizontal average is performed at the disc mid-plane, $z=0$) during a strong burst (orbit number 157) in run 5-OHA-5n; at this time and vertical location, $\langle B_x \rangle_{xy} = -0.021$, $\langle B_y \rangle_{xy} = 0.26$, and $\langle B_z \rangle_{xy} = -0.0045$. The top row shows these fluctuations in the $x$-$y$ plane at $z=0$; non-axisymmetry is readily apparent, with $k_y / k_x > 0$. The bottom row shows the same fluctuations in the $y$-$z$ plane at $x=-1$; the dominant modes appear to be tilted in this plane, with $k_y/k_z > 0$. This mode geometry implies $k_z v_{\rm H} q \Omega_0\simeq k_zk_y\ell_{\rm H}B_y q\Omega_0 > 0$, which exonerates the non-axisymmetric diffusive MRI (see Eq.~\ref{condition}). To test whether the non-axisymmetric HSI, which requires $k_z v_{\rm H} q \Omega_0 > 0$, is responsible for the bursty behaviour, we ask whether $(\oA/\Omega_0)^2 \ll 1$ for the observed fluctuations. For that, we perform two-dimensional Fourier transforms of the panels in Fig.~\ref{bxy_r5}, find the dominant $(k_x,k_y,k_z)$ mode, and use the aforementioned mean-field values to compute $\bm{k}\bcdot\langle\bm{B}\rangle_{xy}$. This procedure gives $(k_x,k_y,k_z) = \pi \times (1,1/4,5/6)$ and so $(\oA/\Omega_0)^2 = 0.016$, which is indeed much smaller than unity. We also check whether these fluctuations satisfy the HSI polarization (Eq.~\ref{hsimode}). Using $k_z v_{\rm H} = 0.50$ and $\chi = 0.40$, as measured for the largest-amplitude fluctuations, in Equation (\ref{hsimode}), we can predict $\delta B_x / \delta B_y = -0.70$. A comparison of this value with the average value measured in Fig.~\ref{bxy_r5}, namely $\delta B_x / \delta B_y = -0.70$, strongly supports our conjecture.

While we have not definitively proven that the non-axisymmetric HSI is responsible for driving the large-amplitude variability seen in some of our simulations, we have shown that the fluctuations observed in these simulations that are associated with this variability have attributes consistent with those expected from this instability. In any case, the important point remains. Under certain conditions pertinent to the inner regions of protoplanetary discs, the Hall effect, acting on a differentially rotating disc that is threaded by a weak magnetic field anti-aligned with the rotation axis, can lead to bursts of turbulence and enhanced angular-momentum transport. These are the first disc simulations to demonstrate this.

Finally, as has been pointed out by several authors \cite[e.g.][]{simon13b,bai13b,lesur13}, outflows can be launched from the disk at the top and bottom of the shearing box, but in opposite radial directions. In this (unphysical) geometry, the toroidal field at any given time does not reverse sign along the $z$ dimension, as it would if the outflows had a physically realistic geometry. When present, the current sheet associated with this sign flip could lead to additional dissipation that would quench the non-axisymmetric HSI modes and prevent the bursts from occurring in the first place. We have examined the toroidal field geometry for the runs that demonstrate bursts and find that while two of the simulations have no toroidal field reversal (5-OHA-4n and 5-OHA-5n), the remaining simulation (10-OHA-5n) does have such a reversal and still undergoes this variable accretion.  Thus, we do not expect that the large scale magnetic field geometry will affect the existence of these bursts.

\begin{figure*}
\begin{center}
\leavevmode
\includegraphics[width=1.9\columnwidth]{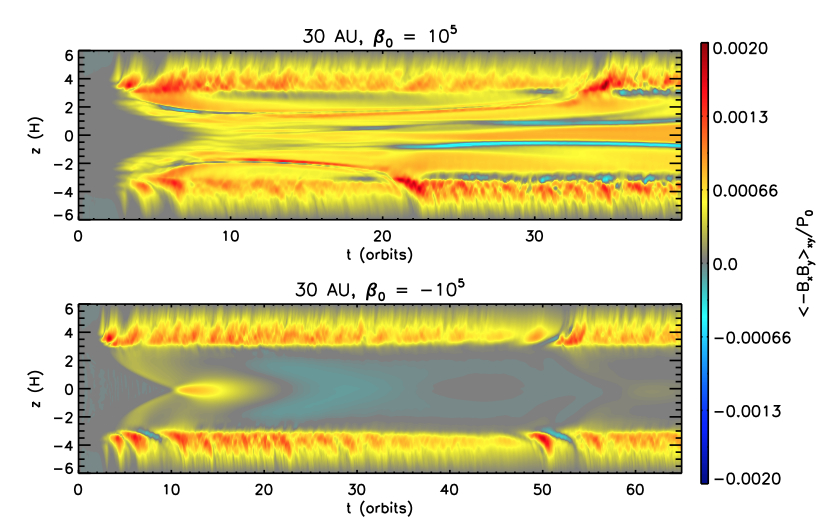}
\end{center}
\caption[]{Space-time diagram of the horizontally averaged Maxwell stress, normalized by the initial, mid-plane gas pressure for runs 30-OHA-5p ($30~{\rm au}$, $\beta_0 = 10^5$; top) and 30-OHA-5n ($30~{\rm au}$, $\beta_0 = -10^5$; bottom).  While both runs show significant Maxwell stress for $|z| \gtrsim 3$, the $\beta_0 < 0$ simulation has very little stress in the mid-plane region while the $\beta_0 > 0$ case exhibits a significant Maxwell stress in this region.}
\label{sttz_max_30au_beta1e5}
\end{figure*}
\begin{figure*}
\centering
\leavevmode
\includegraphics[width=1.9\columnwidth]{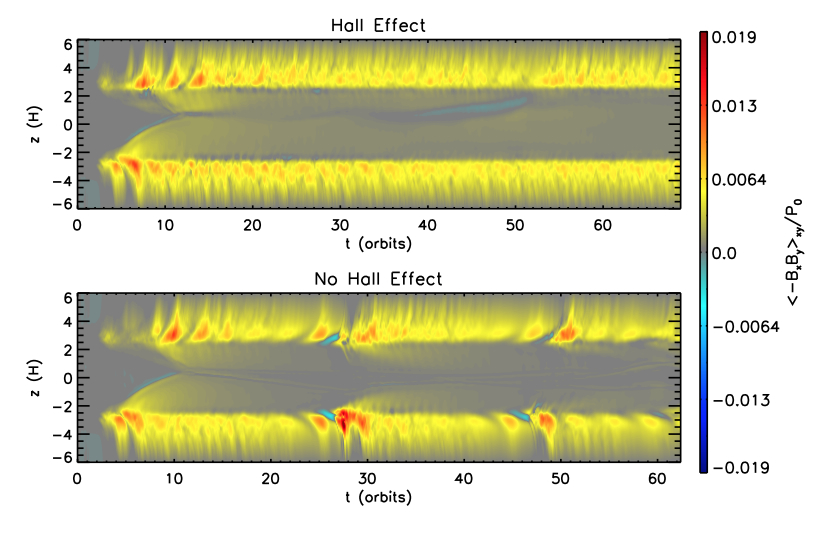}
\caption[]{Space-time diagram of the horizontally averaged Maxwell stress (normalized by the initial, mid-plane gas pressure) for runs 100-OHA-4p-Am1 (top) and 100-OA-4p-Am1 (bottom).  The neglect of the Hall effect in run 100-OA-4p-Am1 does not affect the evolution of the Maxwell stress.}
\label{sttz_max_100au_am1}
\includegraphics[width=1.9\columnwidth]{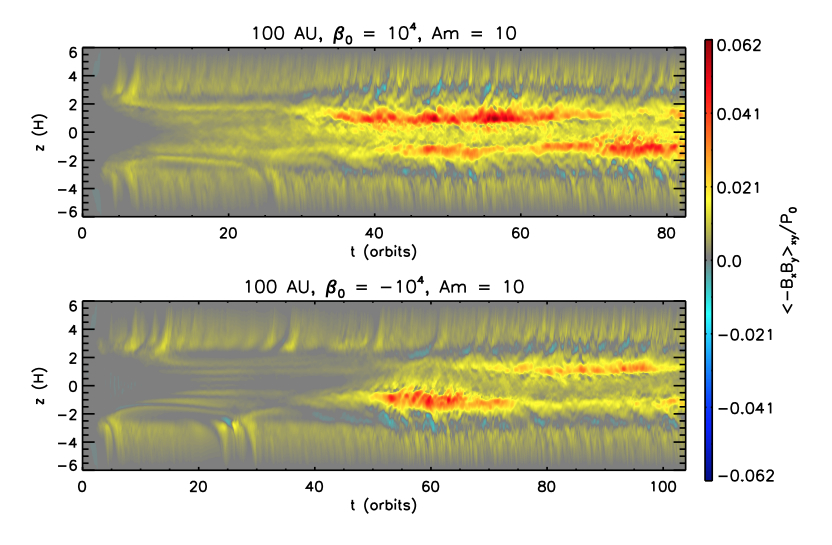}
\caption[]{Space-time diagram of the horizontally averaged Maxwell stress (normalized by the initial, mid-plane gas pressure) for runs 100-OHA-4p ($100~{\rm au}$, $\beta_0 = 10^4$; top) and 100-OHA-4n ($100~{\rm au}$, $\beta_0 = -10^4$; bottom).  Both runs show significant stress near the disc mid-plane, which is independent of the orientation of the initial vertical magnetic field.}
\label{sttz_max_100au_am10}
\end{figure*}

\subsection{Outer Disc}

In this Section, we focus on radii larger than $10~{\rm au}$. Fig.~\ref{sttz_max_30au_beta1e5} shows the space-time evolution of the horizontally averaged $xy$ Maxwell stress for runs 30-OHA-5p and 30-OHA-5n ($\beta_0 = \pm 10^5$). There is very little to no Maxwell stress within the mid-plane region of 30-OHA-5n (i.e.~below the FUV-ionized surface layers, which corresponds to $|z| \lesssim 2H_0$), whereas there is significant stress in this region for 30-OHA-5p.  This result is in agreement with recent  numerical work by \citep{bai15} and with the expectation, borne out of linear theory \citep{wardle99a,balbus01}, for the Hall effect to make the disc more (less) unstable when $\bmath{\Omega} \bcdot {\bmath B} > 0$ ($\bmath{\Omega} \bcdot {\bmath B} < 0$).   Indeed, there is nearly two orders of magnitude difference in $\alpha_{\rm mid}$ between the $\beta_0 = 10^3$ and $\beta_0 = -10^3$ cases.  Other initial magnetic-field strengths (see Table~\ref{tbl:sims}) show similar results, though the differences between $\alpha_{\rm mid}$ for the two field polarities are less dramatic. The total volume-integrated $\alpha$, on the other hand, decreases by just a factor of order unity when the polarity of the net vertical field is reversed.  In summary, the Hall effect affects the mid-plane dynamics significantly at $30~{\rm au}$, but does not have a large impact on the total integrated stress suggesting that the total volume-integrated stress is dominated by the surface contribution, consistent with the strong stress observed at high $|z|$ in Fig.~\ref{sttz_max_30au_beta1e5}.

Fig.~\ref{sttz_max_100au_am1} shows the space-time evolution of the horizontally averaged Maxwell stress from the simulations at $100~{\rm au}$ with ${\rm Am} = 1$ at the mid-plane with (100-HA-4p-Am1) and without (100-A-4p-Am1) the Hall effect. Comparing $\alpha_{\rm mid}$ for these two runs, we see that the mid-plane stress differs by less than a factor of two.  We also examined the effect of increasing the mid-plane value of ${\rm Am}$ from $1$ to $10$ (runs 100-HA-4p-Am10 and 100-HA-4n-Am10). Fig.~\ref{sttz_max_100au_am10} gives a space-time comparison of the Maxwell stress in these runs, which have different magnetic-field polarities. From this plot, and by comparing the values of $\alpha_{\rm mid}$ between 100-HA-4p-Am10 and 100-HA-4n-Am10, we find that the stresses in the mid-plane region are mostly independent of the field orientation. Ambipolar diffusion, not the Hall effect, is largely responsible for the stress at these large radii (consistent with the recent work of \citealt{bai15}).

\subsection{Summary of results}

\begin{figure}
\begin{center}
\leavevmode
\includegraphics[width=\columnwidth]{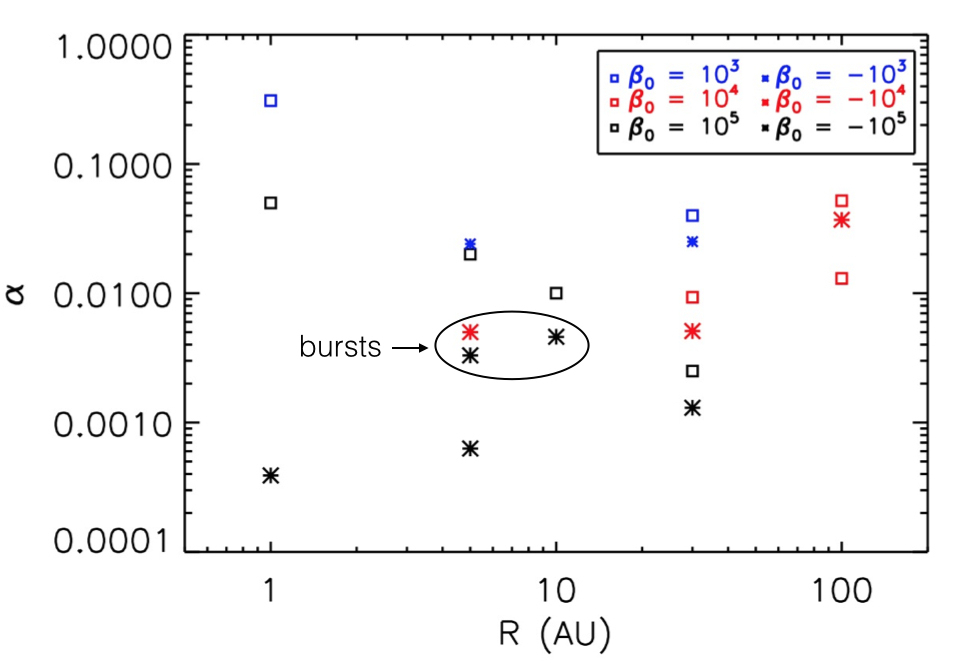}
\end{center}
\caption[]{$\alpha$ versus radius as measured from our shearing-box simulations. The colors denote the magnetic-field strength while the symbols denote polarity, with squares (asterisks) corresponding to $\beta_0 > 0$ ($< 0$).  Multiple values of $\alpha$ at a given radius result from different simulations performed at the same radius but with different diffusion values (see Table~\ref{tbl:sims}).  All runs except for 1-OHA-5n-axi, 5-OA-5n, and 100-A-4p-Am1 are included in this plot. At large radii, the magnetic polarity has minimal effect on the value of $\alpha$, unlike at small radii.  The presence of turbulence `bursts' (circled points) reduces the effect of the magnetic polarity on the amplitude of the stress.}
\label{stuff_radius}
\end{figure}

To summarize Section~\ref{results}, we have plotted $\alpha$ versus $R$ for (nearly) all of our simulations in Fig.~\ref{stuff_radius}.  The strength of angular momentum transport strongly depends on the magnitude and orientation of the magnetic field with respect to the disc angular momentum.  The dependence on the orientation weakens significantly towards large radii and in simulations that undergo `bursts' of turbulence (at $R = 5$--$10~{\rm au}$ ).

\section{Implications}
\label{discussion}

Our results have implications both for theoretical modeling and for future observations.  Here, we discuss these implications and construct an overall scenario for mass accretion in magnetized protoplanetary discs based on our results and those of other recent numerical efforts. 

The most robust inference from current simulations is that, in magnetized protoplanetary discs, a vertical magnetic field threading the disc is necessary to generate rates of angular-momentum transport consistent with observational constraints. This result applies to both the inner and outer discs \citep{bai13b,bai13c,simon13a,simon13b}, and immediately implies that the dependence of the Hall effect on the magnetic-field direction \citep[e.g.][]{wardle99a} might have an observable effect. 
 
The nature of the magnetic stresses responsible for angular-momentum transport varies with radius via changes in the net magnetic-field strength, the chemical properties of the disc, and the degree and nature of the disc ionization. All three of the mechanisms that are possible in principle -- turbulence, laminar stresses internal to the disc, and large-scale wind torques -- are found to be important in simulations. In the inner disc (generally $R \lesssim 10$--$30~{\rm au}$), \citet{bai15} and \citet{lesur14} have emphasized the dominant role of laminar stresses. The MRI can be suppressed at these radii, with the internal laminar MHD stresses arising from the interaction of Keplerian shear and the Hall effect (HSI). Such laminar stresses should be interpreted with caution, as they have so far been seen only in local shearing-box simulations, which are not designed to capture features with horizontal correlation lengths larger than the disc vertical scale height. Those generated by the HSI in our simulations have effectively infinite horizontal correlation lengths, and so the simulations only indicate that the disc is susceptible to generating large-scale global structures. The nature of these structures, and their modeling within the confines of $\alpha$-disc theory, remain largely speculative.

 In the outer disc ($R \gtrsim 30~{\rm au}$), the stresses responsible for accretion can either be turbulent or laminar, depending upon the strength of the magnetic field and the column depth to which FUV photons ionize the gas \citep{simon13a,simon13b}. In the case of turbulent accretion, the turbulence is `layered', similar to the classic dead-zone model of \cite{gammie96}, but with strong ambipolar diffusion near the mid-plane significantly reducing the efficacy of magnetorotational turbulence.  Because this turbulence is not fully quenched in this region, it has been referred to as the `ambipolar damping zone'.

The results presented here have added to and expanded upon this emerging paradigm for disc accretion in magnetized protolanetary discs. One general consideration, which we have emphasized here, is that our ability to accurately model MHD stresses depends upon a knowledge of the relative importance of different non-ideal effects as a function of radius and height within the disc. The general level of agreement between our results and those of \cite{bai15} suggests that the overall scenario outlined above is robust to algorithmic differences in the treatment of the Hall effect. However, within the range of diffusion coefficients informed by chemical models, there are regions of parameter space in which ({\em i}) turbulence can be induced in the inner-disc region if the large-scale magnetic field is anti-aligned with the angular momentum vector of the disc and ({\em ii}) turbulence can be enhanced near the mid-plane in the outer disc due to lower ambipolar diffusion.  Our results thus suggest that discs are not far from the boundary between laminar and turbulent behaviour, and that generally similar discs could fall either way depending on the exact details of the ionization and recombination processes at work.  While improving our modeling of ionization/recombination physics through, e.g., improved chemistry calculations and dust modeling will help us converge on a more robust set of predictions for the behaviour of accreting gas, perhaps the most promising avenue forward is to develop stringent observational tests of our theoretical models and use those tests to constrain the behaviour of observed disc systems. 

\begin{figure*}
\begin{center}
\leavevmode
\includegraphics[width=1.5\columnwidth]{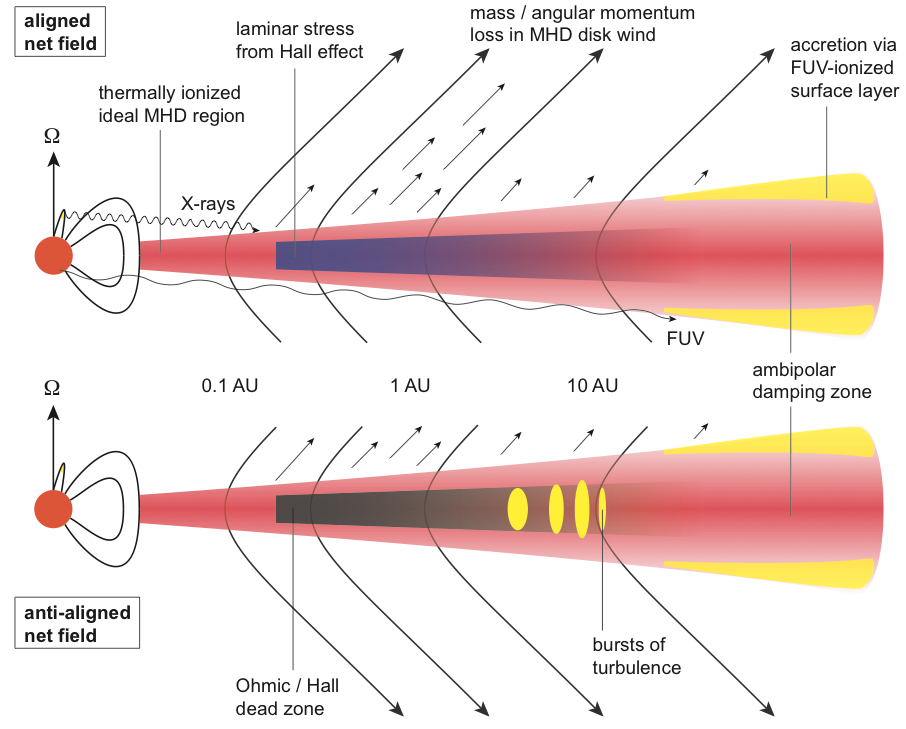}
\end{center}
\caption[]{Illustration of the predicted structure of poorly ionized protoplanetary discs threaded by weak ($\beta_0 \sim \pm 10^{4-5}$) vertical magnetic fields. The structure of the inner, thermally ionized disc, and the outer disc where ambipolar diffusion is dominant, is dependent upon the strength of the net field but is independent of its sign. Between $\sim$$1~{\rm au}$ and $10~{\rm au}$, however, the Hall effect is dominant and the disc structure depends on the sign of $\bm{\Omega} \bcdot \bm{B}$.}
\label{hall_cartoon}
\end{figure*}

Recently, a promising avenue for testing accretion theories has emerged, namely to predict and then observationally constrain signatures of turbulence in protoplanetary discs \citep{carr04,hughes11,simon11b,guilloteau12,forgan12,degregorio-monsalvo13,shi14,simon15a}. As illustrated in Fig.~\ref{hall_cartoon}, our results suggest qualitatively that the magnitude and {\it nature} of angular momentum transport are bi-modal, depending upon the orientation of the net vertical magnetic field with respect to the rotation axis. If $\bm{\Omega} \bcdot \bm{B} > 0$, the stress between $R\sim 1~{\rm au}$ and $R\sim10~{\rm au}$\footnote{The exact radial locations that encompass these different regimes will depend on the surface density model adopted.} is large and mostly laminar; if $\bm{\Omega}\bcdot\bm{B} < 0$, the stress is on average smaller and may exhibit bursty behaviour. The Hall effect, which is responsible for this behaviour, becomes less important farther out in the disc, as we have shown here, consistent with both analytic \citep{salmeron05,salmeron08} and numerical \citep{bai15} expectations. 

\citet{simon15a} carried out numerical simulations focused on the outer regions of protoplanetary discs, which included Ohmic and ambipolar diffusion, in order to predict signatures of turbulence that could be constrained via molecular-line observations.   Our results suggest that the predictions made by \cite{simon15a} are not substantially affected by their neglect of the Hall effect at radii $R \sim 100~{\rm au}$.  Sub-mm observations are therefore predicted to show that the large-scale properties of discs are consistent with a single population. However, observations sensitive to scales smaller than $30~{\rm au}$ could start to reveal the bi-modality produced in Hall-dominated accretion.  Observations in the infrared \citep{carr04} are a potentially better probe of turbulence in the inner disc regions and could thus be a very important tool for testing the predicted bi-modality of turbulent behaviour.


\section{Conclusions}
\label{conclusions}

We have presented results from a series of local stratified shearing-box simulations of magnetorotationally driven protoplanetary disc accretion. The simulations assume a disc model based on the minimum mass solar nebula \citep{hayashi81}, in most cases threaded by a weak flux of vertical magnetic field. Our main conclusions are:
\begin{enumerate}
\item The mid-plane of protoplanetary discs is not necessarily `dead' and may have either turbulent or laminar Maxwell stresses that drive accretion at an enhanced rate ($\alpha \sim 0.001$--$0.1$).
\item In the inner disc ($R_0 \lesssim 30~{\rm au}$), the nature of the mid-plane angular-momentum transport depends upon the orientation of the large-scale vertical magnetic field with respect to the disc rotation axis, $\bm{\Omega} \bcdot \bm{B}$. If the field and rotation vectors are aligned (i.e.~$\bm{\Omega} \bcdot \bm{B} > 0$), angular momentum is transported through a laminar Maxwell stress, in agreement with previous simulations by \citet{lesur14} and \citet{bai15}. If the field is anti-aligned with the rotation axis (i.e.~$\bm{\Omega} \bcdot \bm{B} < 0$), then the mid-plane either may remain quiescent or may undergo `bursts' of turbulent activity, depending upon the precise values of the diffusivities and magnetic-field strength in that region. We presented evidence that these bursts of turbulence are driven by the non-axisymmetric version of the HSI, which results from the interaction of Keplerian shear and non-axisymmetric whistler waves on a predominantly toroidal magnetic field.
\item At intermediate radii, $R_0 \sim 30~{\rm au}$, the sign of $\bm{\Omega} \bcdot \bm{B}$ determines whether or not there is angular-momentum transport within the mid-plane region. However, the height-integrated stress is not strongly dependent upon the magnetic-field orientation -- a result of the FUV ionization layer penetrating closer to the mid-plane at large radii.
\item At large radii, $R_0 \gtrsim 100~{\rm au}$, the orientation of the magnetic field has little influence on the turbulent activity in the disc mid-plane and on the overall stress.  This corroborates the conclusions of \cite{bai15}, who used a different numerical scheme for solving the Hall-MHD induction equation.
\end{enumerate}

A long standing goal of accretion-disc theory is to predict the structure and evolution of the disc from a first-principles understanding of the stresses that drive accretion. Our work does not accomplish that goal, first because we assume a particular radial disc structure and chemistry model in calculating the non-ideal terms and, secondly, because we do not address the transport due to winds, for which global simulations are needed. However, local simulations are now relatively computationally inexpensive, and it is possible to calculate the internal stresses as a function of the known control parameters (e.g.~radius, surface density, external magnetic-field strength, disc chemistry) by carrying out a large number of simulations that sweep through parameter space.  The simulations carried out in this work represent a small, yet informative, region of this parameter space. While much of this parameter space remains to be probed via numerical simulations, our results suggest that any universal prediction to be made for quantities such as $\Sigma(\dot{M},R)$ will be quite difficult to formulate, at least if MHD stresses dominate the evolution.  Instead, quite minor changes to the adopted chemistry model can make a qualitative difference to the angular-momentum transport in some regions of the disc. Moreover, the net magnetic field is a difficult-to-measure quantity, which will likely differ from disc to disc and which will evolve over time in a complex way dependent upon the radial structure of the disc turbulence. An MHD analog of an $\alpha$-model for protoplanetary discs, which predicts the structure given a small number of parameters that could be observationally constrained, does not seem realistically attainable at this time.

Adopting a more optimistic tone, there are excellent prospects for testing whether the MHD processes that we have discussed are actually responsible for the evolution of protoplanetary discs. Our results point toward a basic paradigm of protoplanetary disc evolution that includes both laminar and turbulent accretion, depending upon the strength and orientation of the magnetic field.  The outer disc will be accreting, independent of the orientation of the large-scale vertical magnetic field, through either turbulent or laminar stresses. The details of the accretion dynamics depend upon the strength of the net vertical magnetic field, and may be testable with current sub-millimeter observations \citep{simon15a}. On smaller radial scales, large-scale correlations in the radial and toroidal components of the magnetic field and magnetocentrifugal winds \citep{bai13a,bai13b} are likely to be dominant, although we have emphasized that even here there is the potential for turbulence. Possible observational tests on $\sim$au scales include the detection of an MHD wind or of bi-modality in the nature and amplitude of the angular-momentum transport that is predicted to arise from the Hall effect. 
 
In summary, while the nature of accretion in protoplanetary discs is intimately tied to the details of ionization/recombination physics, thereby making precise numerical predictions difficult, there remain robust, qualitative predictions related to the importance of the large-scale vertical magnetic field in driving accretion and the bi-modality of disc properties that results from the interaction of the Hall effect and this large-scale field.  With powerful new observing facilities, such as ALMA, testing these predictions is now entering the realm of feasibility. 

\section*{Acknowledgements}
We thank Sebastien Fromang, Xue-Ning Bai, and Jim Stone for useful discussions and advice.  We also thank the anonymous referee whose suggestions improved the quality of this manuscript. We acknowledge support from NASA through grants NNX13AI58G (PJA), from the NSF through grant AST 1313021 (PJA), and from grant HST-AR-12814 (PJA) awarded by the Space Telescope Science Institute, which is operated by the  Association of Universities for Research in Astronomy, Inc., for NASA, under contact NAS 5-26555. Support for JBS was provided in part under contract with the California Institute of Technology (Caltech) and the Jet Propulsion Laboratory (JPL) funded by NASA through the Sagan Fellowship Program executed by the NASA Exoplanet Science Institute.  The computations were performed on Stampede and Maverick at the Texas Advanced Computing Center and on Kraken and Darter at the National Institute for Computational Sciences using XSEDE grant TG-AST120062.

\label{lastpage}

\end{document}

%% file: shortcuts.tex
\newcommand{\D}[2]{\frac{\partial #2}{\partial #1}}

\newcommand{\deriv}[2]{\frac{{\rm d} #2}{{\rm d} #1}}

\newcommand\bm[1]{\mbox{\boldmath{$#1$}}}
\newcommand\del{\bm{\nabla}}
\newcommand\bcdot{\bm{\cdot}}
\newcommand\btimes{\bm{\times}}

\newcommand{\ex}{\hat{\bm{e}}_x}
\newcommand{\ey}{\hat{\bm{e}}_y}
\newcommand{\ez}{\hat{\bm{e}}_z}
\newcommand{\alf}{{\rm Alfv\acute{e}n}}
\newcommand{\cs}{c_{\rm s}}  
\newcommand{\va}{v_{\rm A}}
\newcommand{\etao}{\eta_{\rm O}}
\newcommand{\etah}{\eta_{\rm H}}
\newcommand{\etaa}{\eta_{\rm A}}
\newcommand{\rem}{R_{\rm m}}

\newcommand\apj{{ApJ}}%
\newcommand\aap{{A\&A}}%
\newcommand\mnras{{MNRAS}}%
\newcommand\pasj{{PASJ}}%

%% file: ms.bbl
\begin{thebibliography}{64}
\expandafter\ifx\csname natexlab\endcsname\relax\def\natexlab#1{#1}\fi

\bibitem[{Armitage(2011)}]{armitage11}
Armitage P.~J., 2011, Annual Review of Astronomy and Astrophysics, 49, 195

\bibitem[{Bai(2013)}]{bai13c}
Bai X.-N., 2013, The Astrophysical Journal, 772, 96

\bibitem[{Bai(2014)}]{bai14a}
Bai X.-N., 2014, The Astrophysical Journal, 791, 137

\bibitem[{Bai(2015)}]{bai15}
Bai X.-N., 2015, The Astrophysical Journal, 798, 84

\bibitem[{Bai \& Goodman(2009)}]{bai09}
Bai X.-N., Goodman J., 2009, The Astrophysical Journal, 701, 737

\bibitem[{Bai \& Stone(2011)}]{bai11a}
Bai X.-N., Stone J.~M., 2011, The Astrophysical Journal, 736, 144

\bibitem[{Bai \& Stone(2013{\natexlab{a}})}]{bai13a}
Bai X.-N., Stone J.~M., 2013{\natexlab{a}}, The Astrophysical Journal, 767, 30

\bibitem[{Bai \& Stone(2013{\natexlab{b}})}]{bai13b}
Bai X.-N., Stone J.~M., 2013{\natexlab{b}}, The Astrophysical Journal, 769, 76

\bibitem[{Balbus \& Hawley(1991)}]{balbus91}
Balbus S.~A., Hawley J.~F., 1991, ApJ, 376, 214

\bibitem[{Balbus \& Hawley(1992)}]{balbus92}
Balbus S.~A., Hawley J.~F., 1992, ApJ, 400, 610

\bibitem[{Balbus \& Hawley(1998)}]{balbus98}
Balbus S.~A., Hawley J.~F., 1998, Reviews of Modern Physics, 70, 1

\bibitem[{Balbus \& Terquem(2001)}]{balbus01}
Balbus S.~A., Terquem C., 2001, ApJ, 552, 235

\bibitem[{Blaes \& Balbus(1994)}]{blaes94}
Blaes O.~M., Balbus S.~A., 1994, ApJ, 421, 163

\bibitem[{Blandford \& Payne(1982)}]{blandford82}
Blandford R.~D., Payne D.~G., 1982, Monthly Notices of the Royal Astronomical
  Society, 199, 883

\bibitem[{Carr, Tokunaga \& Najita(2004)Carr, Tokunaga, \& Najita}]{carr04}
Carr J.~S., Tokunaga A.~T., Najita J., 2004, The Astrophysical Journal, 603,
  213

\bibitem[{Cleeves, Adams \& Bergin(2013)Cleeves, Adams, \& Bergin}]{cleeves13}
Cleeves L.~I., Adams F.~C., Bergin E.~A., 2013, The Astrophysical Journal, 772,
  5

\bibitem[{de~Gregorio-Monsalvo {et~al}\mbox{.}(2013)de~Gregorio-Monsalvo,
  M{\'e}nard, Dent, Pinte, L{\'o}pez, Klaassen, Hales, Cort{\'e}s, Rawlings,
  Tachihara, Testi, Takahashi, Chapillon, Mathews, Juhasz, Akiyama, Higuchi,
  Saito, Nyman, Phillips, Rod{\'o}n, Corder, \&
  Van~Kempen}]{degregorio-monsalvo13}
de~Gregorio-Monsalvo I. {et~al.}, 2013, Astronomy and Astrophysics, 557, A133

\bibitem[{Desch(2004)}]{desch04}
Desch S.~J., 2004, The Astrophysical Journal, 608, 509

\bibitem[{Evans \& Hawley(1988)}]{evans88}
Evans C.~R., Hawley J.~F., 1988, ApJ, 332, 659

\bibitem[{Forgan, Armitage \& Simon(2012)Forgan, Armitage, \& Simon}]{forgan12}
Forgan D., Armitage P.~J., Simon J.~B., 2012, Monthly Notices of the Royal
  Astronomical Society, 426, 2419

\bibitem[{{Fromang}, {Terquem} \& {Balbus}(2002){Fromang}, {Terquem}, \&
  {Balbus}}]{fromang02}
{Fromang} S., {Terquem} C., {Balbus} S.~A., 2002, \mnras, 329, 18

\bibitem[{Gammie(1996)}]{gammie96}
Gammie C.~F., 1996, ApJ, 457, 355

\bibitem[{Gressel {et~al}\mbox{.}(2015)Gressel, Turner, Nelson, \&
  McNally}]{gressel15}
Gressel O., Turner N.~J., Nelson R.~P., McNally C.~P., 2015, The Astrophysical
  Journal, 801, 84

\bibitem[{Guilloteau {et~al}\mbox{.}(2012)Guilloteau, Dutrey, Wakelam, Hersant,
  Semenov, Chapillon, Henning, \& Pi{\'e}tu}]{guilloteau12}
Guilloteau S., Dutrey A., Wakelam V., Hersant F., Semenov D., Chapillon E.,
  Henning T., Pi{\'e}tu V., 2012, Astronomy and Astrophysics, 548, 70

\bibitem[{Hartmann {et~al}\mbox{.}(1998)Hartmann, Calvet, Gullbring, \&
  D'Alessio}]{hartmann98a}
Hartmann L., Calvet N., Gullbring E., D'Alessio P., 1998, The Astrophysical
  Journal, 495, 385

\bibitem[{Hawley, Gammie \& Balbus(1995)Hawley, Gammie, \& Balbus}]{hawley95a}
Hawley J.~F., Gammie C.~F., Balbus S.~A., 1995, ApJ, 440, 742

\bibitem[{Hawley \& Stone(1998)}]{hawley98}
Hawley J.~F., Stone J.~M., 1998, ApJ, 501, 758

\bibitem[{Hayashi(1981)}]{hayashi81}
Hayashi C., 1981, Progress of Theoretical Physics Supplement, 70, 35

\bibitem[{Hughes {et~al}\mbox{.}(2011)Hughes, Wilner, Andrews, Qi, \&
  Hogerheijde}]{hughes11}
Hughes A.~M., Wilner D.~J., Andrews S.~M., Qi C., Hogerheijde M.~R., 2011, The
  Astrophysical Journal, 727, 85

\bibitem[{Igea \& Glassgold(1999)}]{igea99}
Igea J., Glassgold A.~E., 1999, The Astrophysical Journal, 518, 848

\bibitem[{{Ilgner} \& {Nelson}(2006)}]{ilgner06}
{Ilgner} M., {Nelson} R.~P., 2006, \aap, 445, 205

\bibitem[{Kunz(2008)}]{kunz08}
Kunz M.~W., 2008, Monthly Notices of the Royal Astronomical Society, 385, 1494

\bibitem[{Kunz \& Balbus(2004)}]{kunz04}
Kunz M.~W., Balbus S.~A., 2004, Monthly Notices of the Royal Astronomical
  Society, 348, 355

\bibitem[{Kunz \& Lesur(2013)}]{kunz13}
Kunz M.~W., Lesur G., 2013, Monthly Notices of the Royal Astronomical Society,
  434, 2295

\bibitem[{Kunz \& Mouschovias(2009)}]{kunz09}
Kunz M.~W., Mouschovias T.~C., 2009, The Astrophysical Journal, 693, 1895

\bibitem[{Lesur, Ferreira \& Ogilvie(2013)Lesur, Ferreira, \&
  Ogilvie}]{lesur13}
Lesur G., Ferreira J., Ogilvie G.~I., 2013, Astronomy and Astrophysics, 550, 61

\bibitem[{Lesur, Kunz \& Fromang(2014)Lesur, Kunz, \& Fromang}]{lesur14}
Lesur G., Kunz M.~W., Fromang S., 2014, Astronomy and Astrophysics, 566, 56

\bibitem[{Mignone {et~al}\mbox{.}(2007)Mignone, Bodo, Massaglia, Matsakos,
  Tesileanu, Zanni, \& Ferrari}]{mignone07a}
Mignone A., Bodo G., Massaglia S., Matsakos T., Tesileanu O., Zanni C., Ferrari
  A., 2007, ApJS, 170, 228

\bibitem[{Pandey \& Wardle(2012)}]{pandey12}
Pandey B.~P., Wardle M., 2012, Monthly Notices of the Royal Astronomical
  Society, 423, 222

\bibitem[{Perez-Becker \& Chiang(2011)}]{perez-becker11b}
Perez-Becker D., Chiang E., 2011, The Astrophysical Journal, 735, 8

\bibitem[{{R{\"u}diger} \& {Kitchatinov}(2005)}]{rudiger05}
{R{\"u}diger} G., {Kitchatinov} L.~L., 2005, \aap, 434, 629

\bibitem[{Salmeron, K{\"o}nigl \& Wardle(2007)Salmeron, K{\"o}nigl, \&
  Wardle}]{salmeron07}
Salmeron R., K{\"o}nigl A., Wardle M., 2007, MNRAS, 375, 177

\bibitem[{{Salmeron} \& {Wardle}(2003)}]{salmeron03}
{Salmeron} R., {Wardle} M., 2003, \mnras, 345, 992

\bibitem[{{Salmeron} \& {Wardle}(2005)}]{salmeron05}
{Salmeron} R., {Wardle} M., 2005, \mnras, 361, 45

\bibitem[{{Salmeron} \& {Wardle}(2008)}]{salmeron08}
{Salmeron} R., {Wardle} M., 2008, \mnras, 388, 1223

\bibitem[{{Sano} {et~al}\mbox{.}(2000){Sano}, {Miyama}, {Umebayashi}, \&
  {Nakano}}]{sano00}
{Sano} T., {Miyama} S.~M., {Umebayashi} T., {Nakano} T., 2000, \apj, 543, 486

\bibitem[{Sano \& Stone(2002{\natexlab{a}})}]{sano02a}
Sano T., Stone J.~M., 2002{\natexlab{a}}, ApJ, 570, 314

\bibitem[{Sano \& Stone(2002{\natexlab{b}})}]{sano02b}
Sano T., Stone J.~M., 2002{\natexlab{b}}, ApJ, 577, 534

\bibitem[{Shakura \& Syunyaev(1973)}]{shakura73}
Shakura N.~I., Syunyaev R.~A., 1973, A\&A, 24, 337

\bibitem[{Shi \& Chiang(2014)}]{shi14}
Shi J.-M., Chiang E., 2014, The Astrophysical Journal, 789, 34

\bibitem[{Shu {et~al}\mbox{.}(1994)Shu, Najita, Ostriker, Wilkin, Ruden, \&
  Lizano}]{shu94}
Shu F., Najita J., Ostriker E., Wilkin F., Ruden S., Lizano S., 1994, The
  Astrophysical Journal, 429, 781

\bibitem[{Simon, Armitage \& Beckwith(2011)Simon, Armitage, \&
  Beckwith}]{simon11b}
Simon J.~B., Armitage P.~J., Beckwith K., 2011, The Astrophysical Journal, 743,
  17

\bibitem[{{Simon} {et~al}\mbox{.}(2013{\natexlab{a}}){Simon}, {Bai},
  {Armitage}, {Stone}, \& {Beckwith}}]{simon13b}
{Simon} J.~B., {Bai} X.-N., {Armitage} P.~J., {Stone} J.~M., {Beckwith} K.,
  2013{\natexlab{a}}, \apj, 775, 73

\bibitem[{{Simon} {et~al}\mbox{.}(2013{\natexlab{b}}){Simon}, {Bai}, {Stone},
  {Armitage}, \& {Beckwith}}]{simon13a}
{Simon} J.~B., {Bai} X.-N., {Stone} J.~M., {Armitage} P.~J., {Beckwith} K.,
  2013{\natexlab{b}}, \apj, 764, 66

\bibitem[{Simon {et~al}\mbox{.}(2015)Simon, Hughes, Flaherty, Bai, \&
  Armitage}]{simon15a}
Simon J.~B., Hughes A.~M., Flaherty K.~M., Bai X.-N., Armitage P.~J., 2015,
  arXiv.org

\bibitem[{Suzuki \& Inutsuka(2009)}]{suzuki09}
Suzuki T.~K., Inutsuka S.-I., 2009, ApJ, 691, L49

\bibitem[{{Turner} {et~al}\mbox{.}(2014){Turner}, {Fromang}, {Gammie}, {Klahr},
  {Lesur}, {Wardle}, \& {Bai}}]{turner14}
{Turner} N.~J., {Fromang} S., {Gammie} C., {Klahr} H., {Lesur} G., {Wardle} M.,
  {Bai} X.-N., 2014, Protostars and Planets VI, 411

\bibitem[{{Umebayashi} \& {Nakano}(1980)}]{umebayashi80}
{Umebayashi} T., {Nakano} T., 1980, \pasj, 32, 405

\bibitem[{Umebayashi \& Nakano(2009)}]{umebayashi09}
Umebayashi T., Nakano T., 2009, The Astrophysical Journal, 690, 69

\bibitem[{Wardle(1999)}]{wardle99a}
Wardle M., 1999, MNRAS, 307, 849

\bibitem[{Wardle(2007)}]{wardle07}
Wardle M., 2007, Astrophysics and Space Science, 311, 35

\bibitem[{Wardle \& Koenigl(1993)}]{wardle93}
Wardle M., Koenigl A., 1993, Astrophysical Journal, 410, 218

\bibitem[{Wardle \& Ng(1999)}]{wardle99b}
Wardle M., Ng C., 1999, Monthly Notices of the Royal Astronomical Society, 303,
  239

\bibitem[{Wardle \& Salmeron(2012)}]{wardle12}
Wardle M., Salmeron R., 2012, Monthly Notices of the Royal Astronomical
  Society, 422, 2737

\end{thebibliography}
